\documentclass[final]{raa}
\usepackage{graphicx,times}
\usepackage{natbib}
\usepackage{amssymb,amsmath}
\usepackage{multirow}
\usepackage{threeparttable}
\usepackage{hyperref}
\usepackage{lineno}   %line number
\hypersetup{pdftitle = The title of my PDF, pdfauthor = My name, pdfsubject= The subject, pdfkeywords = keyword1 keyword2 keyword3}
\hypersetup{colorlinks = true, linkcolor = green, anchorcolor = red, citecolor = blue, filecolor = red, urlcolor = red}

\begin{document}
	
	\title{Spectral Deconvolution Analysis on Olivine-Orthopyroxene Mixtures with Simulated Space Weathering Modifications}
	
	\volnopage{ {\bf 20XX} Vol.\ {\bf X} No. {\bf XX}, 000--000}
	\setcounter{page}{1}
	
	\author{Hui-Jie Han\inst{1}, Xiao-Ping Lu\inst{1 *}, Te Jiang\inst{2}, Chih-Hao Hsia\inst{1} \inst{,3}, Ya-Zhou Yang\inst{4}, Peng-Fei Zhang\inst{1}, Hao Zhang\inst{2} \inst{,5} 
	}
	
	\institute{State Key Laboratory of Lunar and Planetary Sciences, Macau University of Science and Technology, Taipa 999078, Macau, PR China; {\it xplu@must.edu.mo}\\
		\and
		Planetary Science Institute, School of Earth Sciences, China University of Geosciences, Wuhan 430074, China\\
		\and
		CNSA Macau Center for Space Exploration and Science, Taipa 999078, Macau, PR China\\
		\and
		State Key Laboratory of Space Weather, National Space Science Center, Chinese Academy of Sciences, Beijing 100190, China\\
		\and
		CAS Center for Excellence in Comparative Planetology, Hefei 230026, China\\
		\vs \no
		%   {\small Received  ; accepted  }
	}
	
%	\linenumbers

	\abstract{Olivine and pyroxene are important mineral end-members for studying the surface material compositions of mafic bodies. The profiles of visible and near-infrared spectra of olivine-orthopyroxene mixtures systematically varied with their composition ratios. 
		In our experiments, we combine the RELAB spectral database with a new spectral data obtained from some assembled olivine-orthopyroxene mixtures. We found that the commonly-used band area ratio (BAR, \citealt{cloutis1986calibrations}) does not work well on our newly obtained spectral data. To investigate this issue, an empirical procedure based on fitted results by  modified Gaussian model is proposed to analyze the spectral curves. Following  the new empirical procedure, the end-member abundances can be estimated with a 15\% accuracy with some prior mineral absorption features. In addition, the  mixture samples configured in our experiments are also irradiated by pulsed lasers to simulate and investigate the space weathering effects. Spectral deconvolution results confirm that low-content olivine on celestial bodies are difficult to measure and estimate. Therefore, the olivine abundance of space weathered materials may be underestimated from remote sensing data. This study may be used to  quantify the spectral relationship of olivine-orthopyroxene mixtures and further reveal their correlation between the spectra of ordinary chondrites and silicate asteroids.
	}
	
	\authorrunning{H.-J. Han et al. }            %author_head in even pages
	\titlerunning{Spectral Deconvolution on OLV-OPX Mixtures}  % title_head in odd pages
	\maketitle

	%________________________________________________ sections below
	%
	\section{Introductions}           %% first-level sections will be auto-capitalized
	\label{sect:Introductions}
	
	The spectral absorption features provided by hyperspectral remote sensing enable us to investigate the geological and geochemical information over the imaged area (\citealt{cloutis1996review, clark2003imaging, zaini2014determination}). Therefore it is widely used to identify the minerals, rocks, water (ices), and organic materials of the celestial bodies (\citealt{adams1974visible, Singer1981Near, burbine2002spectra, pieters2009moon,basilevsky2012geologic,Roush2015Laboratory,lauretta2019unexpected}), such as terrestrial planets, moon and asteroids (\citealt{cloutis2000diaspores, ohtake2009global,lindsay2015composition, zhang2015situ, hu2019mineral, tsuda2020hayabusa2}). Visible and near-infrared (VNIR) reflectance spectra are susceptible to the mafic minerals (\citealt{Pieters1988Exploration, lucey2004mineral, Staid2011The}), which contain the characteristics of mineral composition and crystal structure (\citealt{hunt1977spectral, mustard2005olivine}). As the major group of mafic minerals, olivine and pyroxene are important materials in understanding the geologic evolution, differentiation process and  cooling history of terrestrial bodies (\citealt{wells1977pyroxene, ishii1983petrological, cloutis1991pyroxene, klima2011new}). However, many limited factors should be considered when using reflectance spectra to analyze remote sensing data from airless bodies, because most reflectance spectra collected under laboratory are not consistent with the data by remote sensing (\citealt{bishop1998spectroscopic, cloutis2008spectral, shirley2019particle}). One thing that needs to be focused on is “space weathering”, which can change the chemical compositions and optical properties of the  surface materials (\citealt{pieters2000space, brunetto2006space, fu2012effects}). It can make the spectral curves redden and darken, and difficult to identify and interpret (\citealt{hapke2001space,chapman2004space, han2020study}). Therefore, many methods for space weathering spectral deconvolution are  extensively studied in order to acquire the truth data interpretation (\citealt{Sunshine1993Estimating, Sunshine1998Determining, gallie2008equivalence, han2020study, potin2020model}).

	\subsection{Olivine, Pyroxene and Their Mixture Spectroscopy}
	\label{subsec:Olivine, Pyroxene and Their Mixture Spectroscopy}
	
\begin{figure}
		\centering
		\includegraphics[width=12.0cm, angle=0]{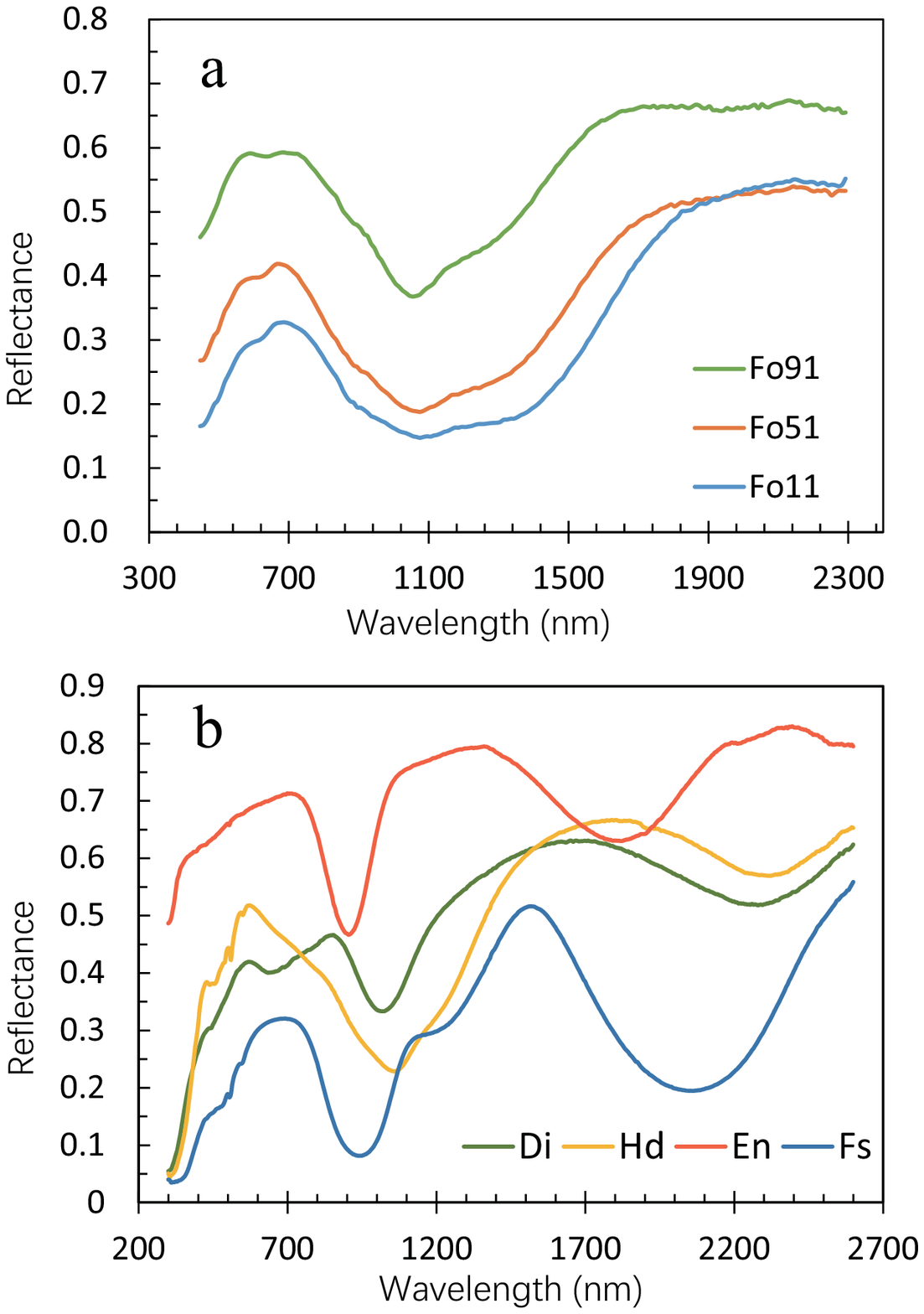}
		\caption{Reflectance spectra of olivine with varied Mg-number (Fo\#) and the ones with pyroxene quadrilateral end-members. (a). Spectra of olivine with varied Fo\#. (b). The comparison of the spectra with various pyroxene quadrilateral end-members: Di (diopside; shown as green), Hd (hedenbergite; orange), En (Enstatite; red) and Fs (Ferrosillite; blue)}.
		\label{Fig1}
	\end{figure}

	Olivine is typically the dominant and foremost mineral crystallizing from liquid magma in the Earth’s mantle and bears rich information about the magma ocean (\citealt{elthon1979high, garcia1995olivine, danyushevsky2000re, dyar2009spectroscopic}). Olivines are mafic silicates with the formula (Mg, Fe)$_{2}$SiO$_{4}$, solid solutions ranging from fayalite (Fe$_{2}$SiO$_{4}$) to forsterite (Mg$_{2}$SiO$_{4}$). The olivine structure is nearly hexagonally close-packed and there are two octahedral sites, M1 and M2. Half of the two octahedral interstices occupied by divalent metal ions, usually either Mg$^{2+}$ or Fe$^{2+}$ (\citealt{dyar2009spectroscopic}). The color of olivine is mainly determined by the content of  Fe$^{2+}$ content. It looks green with lower content of Fe$^{2+}$ but becomes brown at high iron content. In VNIR reflectance spectra, olivine is characterized by its Fe$^{2+}$ electronic transition absorption bands (\citealt{burns1970crystal, hunt1977spectral}), and forming a wide and superimposed absorption near 1.0 $\mu$m region (Figure \ref {Fig1}a). As the Fe$^{2+}$ abundance increases, the 1.0 $\mu$m absorption band center shifts toward longer wavelength and the band width increses (\citealt{Sunshine1998Determining, han2020study}).

	%---------------------------Figures 1----------------------------
	%-----------------------------------------------------------------

	Pyroxenes are crystallized from original magma and forming the primary component of planetary upper mantle and crust, and bear rich magma information (\citealt{elkins2005possible, elardo2011lunar, head2017generation}). Several geothermobarometric tools have focused on pyroxene to study the igneous differentiation in the Earth, asteroids and meteorites (\citealt{lindsley1983pyroxene, cloutis1991pyroxene, Sunshine2004High, klima2008characterization}). Pyroxenes are single-chain silicates with the general formula XYSi$_{2}$O$_{6}$, where X and Y are both divalent cations (commonly Ca, Fe, Mg, etc.). The stacking of chains gives rise to two types of cavities, labelled M1 (close to octahedral in shape) and M2 (more irregular in shape) (\citealt{burns1970crystal}). Fe$^{2+}$ in the two different crystallographic positions M1 and M2 are responsible for the dominant pyroxene spectra absorptions in the VNIR region, which produce two prominent absorption features near 1.0 $\mu$m (Band 1) and 2.0 $\mu$m (Band 2) bands. While substitutions of other cations such as Ca$^{2+}$ change the crystal structure and make the absorption characteristic to move towards longer wavelength (\citealt{adams1974visible, klima2011new}). Furthermore, it was found that in the spectra of the pyroxenes (Figure \ref {Fig1}b), the wavelengths of the two main absorption bands change as a function of the concentrations of Fe$^{2+}$ and Ca$^{2+}$  (\citealt{adams1974visible, klima2007spectroscopy}).

	The spectral properties of pyroxenes have been intensively studied in the past (\citealt{cloutis1991pyroxene, Sunshine1993Estimating}). The most basic division of pyroxenes is between orthopyroxenes (OPX) and clinopyroxenes (CPX) which is based on differences in chemical composition and crystal structure. According to this scheme, orthopyroxenes essentially contains $<$5 mol\% CaSiO$_{3}$ (wollastonite, Wo) and possess orthorhombic symmetry, while clinopyroxenes can contain 0 $\sim$ 50 mol\% Wo and possess monoclinic symmetry (\citealt{adams1974visible, cloutis1991pyroxene}). Therefore, they can form a diagram of pyroxenes quadrilateral, the corners of the quadrilateral are the pyroxene end-members diopside (Di, CaMgSi$_{2}$O$_{6}$), hedenbergite (Hd, CaFeSi$_{2}$O$_{6}$), Enstatite (En, Mg$_{2}$Si$_{2}$O$_{6}$), and Ferrosilite (Fs, Fe$_{2}$Si$_{2}$O$_{6}$). Example spectra are shown in Figure \ref {Fig1}b. The olivines and pyroxenes spectral data are all taken from the Reflectance Experiment Laboratory (RELAB) database support by Brown University. $\footnote{\url{http://www.planetary.brown.edu/relab/}}$.

	According to the information described above, retrieving the mafic mineral features from the obtained reflectance spectra does seem to be a viable task. It provides an effective method for remote analysis of mafic minerals on the planet's surface. However, it is still difficult to directly interpret the spectra because spectral analysis are hampered by the overlapping absorption features, especially when dealing with mixture minerals (\citealt{clenet2011new}). The spectral properties of mineral mixtures do not linearly vary with the relative abundances of different end-members (\citealt{adams1974visible, Singer1981Near,cloutis1991pyroxene}).

\begin{figure}
		\centering
		\includegraphics[width=12.0cm, angle=0]{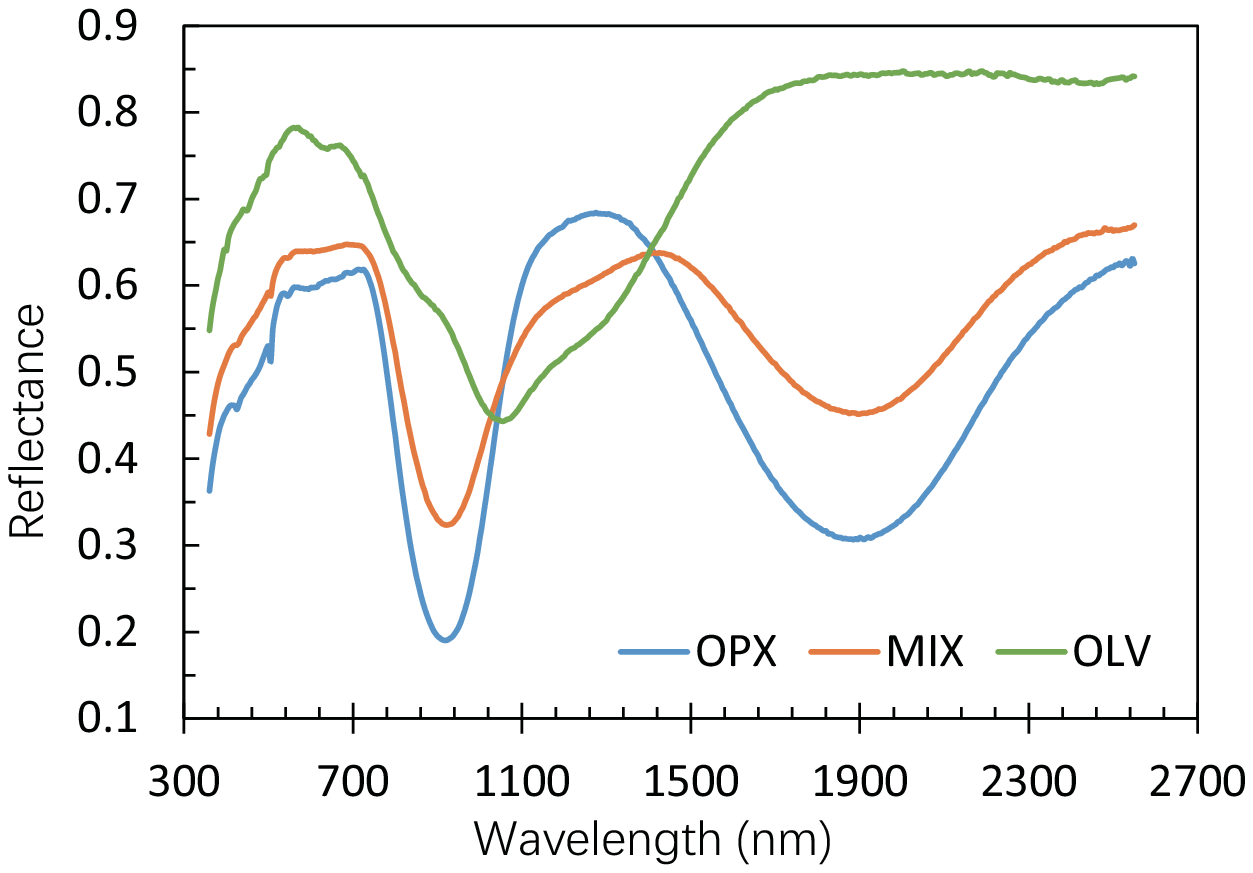}
		\caption{VNIR reflectance spectra of three mineral components with orthopyroxene (shown as blue), olivine (green), and olivine-orthopyroxene mixture (red). The RELAB IDs of these spectra are AG-TJM-008 (OLV), AG-TJM-009 (OPX) and AG-TJM-019 (MIX) respectively.}
		\label{Fig2}
	\end{figure}

	In this study, we choose the forsterite and low-Fe orthopyroxene as the two end-members. Low-Fe orthopyroxene spectra show a narrow strong absorption near 0.92 $\mu$m region and a wide absorption near 1.8 to 1.9 $\mu$m region (\citealt{klima2007spectroscopy}). Therefore, center position of these two absorption bands will be important information for studying the olivine-orthopyroxene mixture spectra. Figure \ref {Fig2} shows the VNIR reflectance spectra of olivine, orthopyroxene and their mixture (50\% OLV + 50\% OPX).
	
	%---------------------------Figures 2----------------------------
	%-----------------------------------------------------------------

	\subsection{Space Weathering Simulations}
	\label{subsec:Space Weathering Simulations}
	
	The spectra of asteroids and lunar regolith are obviously different from those of the laboratory minerals. The former are characterized by distinct red continua slope and weak absorption unlike those of the fresh rocks in the laboratory. Nanophase iron particles (npFe$^{0}$) have been widely observed in lunar soil which is confirmed to be the cause of the difference in these spectra. It is widely thought that the formation of npFe$^{0}$ in lunar soil is the result of space weathering, which include micrometeorite bombardment, energetic particles, as well as cosmic rays (\citealt{dybwad1971radiation, hapke1973darkening, Pieters1993Optical}). To understand the physical and chemical mechanisms of space weathering, many simulation experiments were made in the laboratory (e.g., \citealt{keller1993discovery, yamada1999simulation, hapke2001space,loeffler2009irradiation, fu2012effects, jiang2019bi}). These experiments irradiated pulsed laser beam onto the different samples to simulate the irradiation of micrometeorite bombardments; or implanted proton particles (such as H$^{+}$, He$^{+}$ and Ar$^{+}$) onto the samples to simulate the solar wind implantation; or heated the samples with microwave furnace at high temperature in a vacuum to simulate the space environment. Among these methods, pulsed laser irradiation is effective in generating space weathering effects. Since the pulsed laser can pump energy in a very short time scale, it can be analogous to a high-speed and low-mass dust particle micrometeorite impact (\citealt{yamada1999simulation, Yang2017Optical}). In this study, we utilized the pulsed laser irradiation for space weathering simulations.

	Overview of this study, firstly we collected and assembled some olivine-orthopyroxene mixture spectra, the assembled samples are also irradiated with pulsed laser in our experiment. Some calculation methods are utilized to study these spectra. These spectral data, experiments and research methods will be presented in Section 2. While, these methods have limitations and do not work well for some samples. So we developed an empirical procedure to deal with these spectral data. The results and some discussions are presented in Section 3. The conclusions are summarized in Section 4.

	\section{Samples and Experimental Methods}
	\label{sect:Samples and Experimental Methods}
	
	\subsection{Samples and Reflectance Spectra Measurements}
	\label{subsec:Samples and Reflectance Spectra Measurements}
	
\begin{figure}
		\centering
		\includegraphics[width=11.5cm, angle=0]{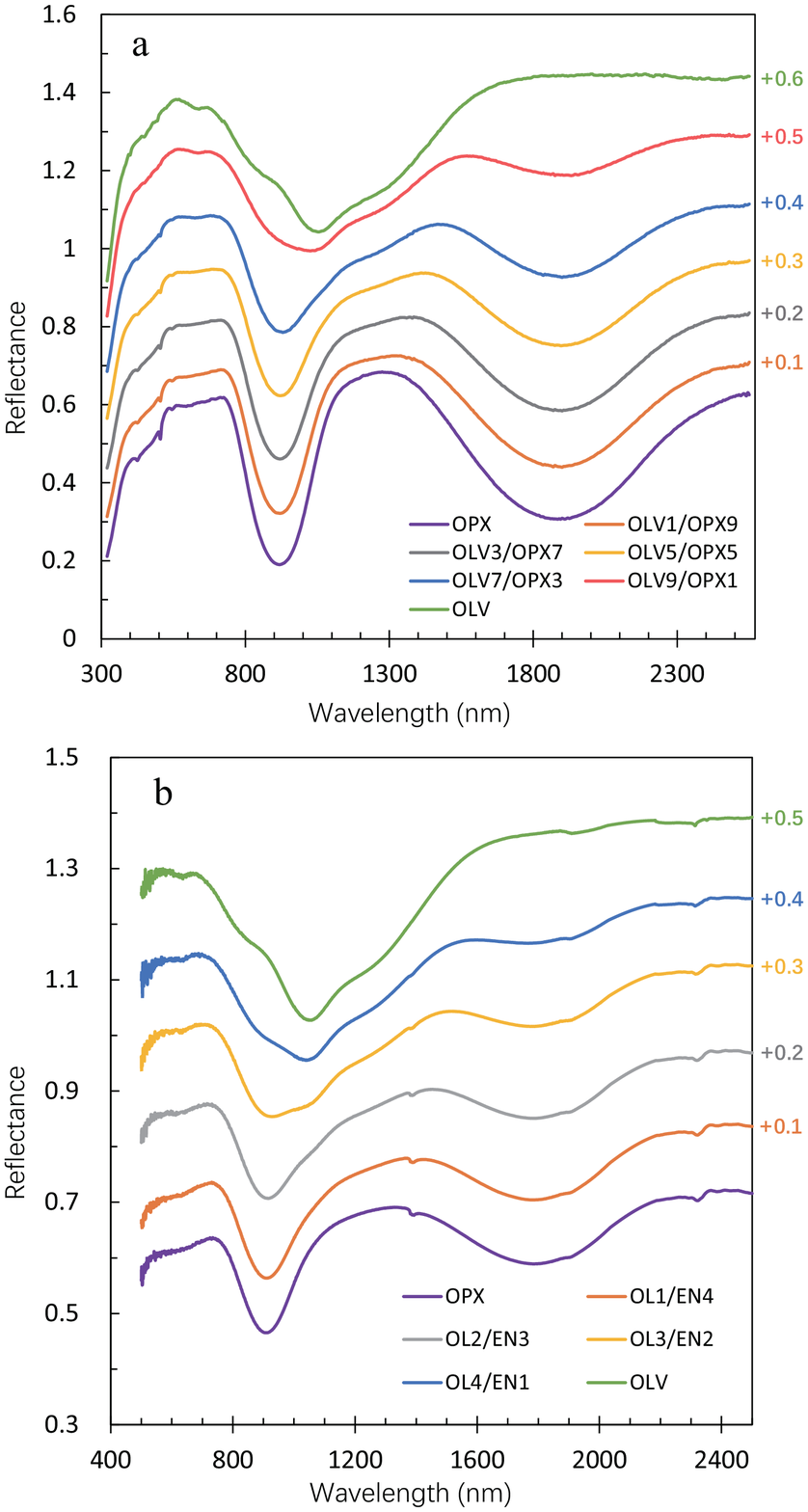}
		\caption{VNIR reflectance spectra of  mixtures composed of olivine and orthopyroxene mixed with different proportions. (a). The spectra are obtained from RELAB database. (b). Spectral profiles are taken from our experiments (2.14 $\mu$m artificial peaks are removed, narrow $\sim$1.41 and $\sim$2.33 $\mu$m absorption bands presented in some spectra are due to minor alteration). For the sake of clarity, with the increase of olivine content, the reflection spectra are superimposed and moved multiple +0.1 on the ordinate.}
		\label{Fig3}
	\end{figure}

	The spectra of synthetic mixtures presented in this study are taken from the RELAB database and our experiment. The end-member components of RELAB synthetic mixture samples are natural olivine (AG-TJM-008) and hypersthene (AG-TJM-009) with grain sizes smaller than 38 $\mu$m. The mass mixing ratio of the two minerals are 1/9, 3/7, 5/5, 7/3 and 9/1 respectively. Another part of synthetic mixture samples from RELAB are natural olivine (PO-CMP-017) and enstatite (PE-CMO-012) mixtures and the grain size is 45 $\sim$ 75 $\mu$m. The mass mixing ratio of the two minerals are 1/9, 2.5/7.5, 5/5, 7.5/2.5 and 9/1 respectively. Chemical compositions of these end-member minerals are listed in Table \ref {Tab1}. The reflectance spectra of olivine-hyperthene mixtures are shown in Figure \ref {Fig3}a. In addition, we also utilize some olivine-bronzite mixture samples whose end-member components are not analyzed. The sample IDs and mixture compositions of these samples are listed in Table \ref {Tab2}. Measurements of these olivine-orthopyroxene mixtures are conducted using a bidirectional VNIR spectrometer with an incidence angle of 30° and an emission angle of 0° ranging from 0.30 (or 0.32) to 2.55 (or 2.60) $\mu$m.

	%------------------------Table  1----------------------------------
	\begin{table}
		\bc
		\begin{minipage}[]{150mm}
			\caption[]{Chemical compositions of these olivine and orthopyroxene samples
				\label{Tab1}}\end{minipage}
		\setlength{\tabcolsep}{2.0mm}
		\small
		\begin{tabular}{ccccccc}
			\hline\noalign{\smallskip}
			Sample ID&  AG-TJM-008 &AG-TJM-009&PO-CMP-017 	&PE-CMP-012  &OWN-OLV&OWN-OPX\\
			\cline{1-7}
			Mineral&Olivine&Hypersthene&	Olivine&	Enstatite&	Olivine&	Enstatite\\
			\cline{1-7}
			SiO$_{2}$&	40.81&	54.09&	40.87&	55.3&	40.66&	52.06\\
			TiO$_{2}$&	0&	0.16&	0&	0.05&	0.011&	0.13\\
			Al$_{2}$O$_{3}$&	0&	1.23&	0.01&	0.12&	0.34&	4.1\\
			Cr$_{2}$O$_{3}$&	0&	0.75&	0.04&	0&	0&	0\\
			Fe$_{2}$O$_{3}$&	0&	0&	0&	0&	9.76&	7.44\\
			FeO&	9.55&	15.22&	7.77&	9.38&	0&	0\\
			MnO&	0.14&	0.49&	0.14&	0.15&	0.11&	0.13\\
			MgO&	49.42&	26.79&	51.58&	32.8&	48.00&	34.85\\
			CaO&	0.05&	1.52&	0.03&	0.45&	0.056&	0.98\\
			Na$_{2}$O&	0&	0.05&	0.01&	0&	0.008&	0.13\\
			K$_{2}$O&	0&	0.05&	0&	0.02&	0.004&	0.006\\
			P$_{2}$O$_{3}$&	0&	0&	0&	0.01&	0.005&	0.016\\
			H$_{2}$O&	0&	0&	0&	0&	0.04&	0.04\\
			\cline{1-7}
			&Fo 90.5&En 73.6&Fo 92.2&	En 85.5&Fo 89.8&En 87.7\\
			Text&	Fa 9.5&	Fs 23.4&	Fa 7.8&	Fs 13.7&	Fa 10.2&	Fs 10.5\\
			& &Wo 3.0& &		Wo 0.8& &		Wo 1.8\\
			\noalign{\smallskip}\hline
		\end{tabular}
		
		\ec
		\tablecomments{0.95\textwidth}{The number unit is wt. \%.}
	\end{table}
	%----------------------------------------------------------------

	%------------------------Table  2----------------------------------
	\begin{table}
		\bc
		\begin{minipage}[]{150mm}
			\caption[]{Information summary of the end-member and synthetic mixed samples
				\label{Tab2}}\end{minipage}
		\setlength{\tabcolsep}{2.0mm}
		\small
		\begin{tabular}{cccccccc}
			\hline\noalign{\smallskip}
			Sample ID&  OLV : OPX & &Sample ID&OLV : OPX &	&Sample ID  &OLV : OPX (Irradiate)\\
			\cline{1-2} \cline{4-5} \cline{7-8}
			AG-TJM-008&	Olivine&&		AG-TJM-017&	1 : 9&&		XT-TXH-030-P&	2.5 : 7.5 (No)\\
			AG-TJM-009&	Hypersthene&&	AG-TJM-018&	3 : 7&&		XT-TXH-031-P&	5 : 5 (No)\\
			PO-CMP-017&	Olivine&&		AG-TJM-019&	5 : 5&&		XT-TXH-032-P&	7.5 : 2.5 (No)\\
			PE-CMP-012&	Enstatite&&		AG-TJM-014&	7 : 3&&		XT-TXH-030-P1&	2.5 : 7.5 (15 mJ)\\
			OWN-OLV&	Olivine&&		AG-TJM-020&	9 : 1&&		XT-TXH-031-P1&	5 : 5 (15 mJ)\\
			OWN-OPX&	Enstatite&&		XO-CMP-015&	9 : 1&&		XT-TXH-032-P1&	7.5 : 2.5 (15 mJ)\\
			OWN-OL1/EN4&	2 : 8&&		XO-CMP-016&	7.5 : 2.5&&	XT-TXH-030-P2&	2.5 : 7.5 (15 mJ * 2)\\
			OWN-OL2/EN3&	4 : 6&&		XO-CMP-017&	5 : 5&&		XT-TXH-031-P2&	5 : 5 (15 mJ * 2)\\
			OWN-OL3/EN2&	6 : 4&&		XO-CMP-018&	2.5 : 7.5&&	XT-TXH-032-P2&	7.5 : 2.5 (15 mJ * 2)\\
			OWN-OL4/EN1&	8 : 2&&		XO-CMP-019&	1 : 9& &	&   \\		
			\noalign{\smallskip}\hline
		\end{tabular}
		\ec
		\tablecomments{0.95\textwidth}{Six samples with IDs like OWN-xxxx come from our experiment, with grain size  less than 75 $\mu$m. The others come from RELAB. Seven samples with IDs like AG-TJM-xxx, with grain size $<$ 38 $\mu$m. The grain size of the other RELAB samples is 45 $\sim$ 75 $\mu$m. Five samples with IDs like XO-CMP-xxx are olivine-enstatite mixture samples. Nine samples with IDs like XT-TXH-xxx are olivine-bronzite mixture samples, and six of these nine samples are irradiated with pulsed laser to study space weathering effect. The total amount of each mixture is counted as 10.}
	\end{table}
	%----------------------------------------------------------------

	%---------------------------Figures 3----------------------------
	%-----------------------------------------------------------------

	The other part of the samples come from our experiment. We collect some natural pure olivine (OWN-OLV) and enstatite (OWN-OPX) samples as experimental materials. The samples are ground and sieved into grain size less than 75 $\mu$m. Chemical compositions of these sample powders are analyzed with wet chemistry method and listed in Table \ref {Tab1}. The mass mixing ratios of the two minerals are 1/4, 2/3, 3/2 and 4/1 respectively (Table \ref {Tab2}). A Bruker Vertex 70 Fourier transform spectrometer is utilized to get the VNIR reflectance spectra of these samples. Each sample is measured four times with the sample holder rotated 90° every time. The average spectrum is taken as the final spectrum to reduce the effect of any surface inhomogeneity (\citealt{jiang2019bi}). What needs our attention here is that the VNIR reflectance spectra of these materials have an artificial peak near 2.14 $\mu$m when Spectralon panel made of polytetrafluoroethylene is used as the calibration plate in the near-infrared (NIR) region (\citealt{Zhang2014Effects}), two narrow spectral faint wobbles near 1.41 and 2.33 $\mu$m absorption bands presented in some spectra are due to minor alteration. These two small fluctuations do not affect the results of this study, because these artificial wobbles are not on the main band absorption regions (Figure \ref {Fig3}b). However, we do need to remove the 2.14 $\mu$m artificial peak, as it appears in the orthopyroxene absorption features. Here, we utilize a 4th-order polynomial to fit the data between 1.8 to 2.5 $\mu$m. Then the data of the artificial peak ($\sim$2.03 to $\sim$2.18 $\mu$m) are replaced for further study. The modified reflectance spectra are shown in Figure \ref {Fig3}b.

	\subsection{Pulsed Laser Irradiation}
	\label{subsec:Pulsed Laser Irradiation}
	
	The detailed experiment process can be found in \cite{jiang2019bi}. Here is a brief summary: In order to study the space weathering effects on olivine, orthopyroxene and their mixture spectra, some natural pure olivine and orthopyroxene powders (less than 75 $\mu$m) are mixed with a certain fractions. Each time $\sim$1.0 g of powder samples are uniformly placed in a 2.5 cm $\times$ 2.5 cm square in an aluminum holder. These powder samples are baked in a dry oven at 120 ℃ for about 12 hours to remove the moisture. The dried samples covered by a thin glass slide to prevent the powders sputtering are placed in a vacuum chamber to cool to room temperature and then subjected to pulsed laser irradiation at a pressure of 10$^{-3}$ Pa.

\begin{figure}
		\centering
		\includegraphics[width=12.0cm, angle=0]{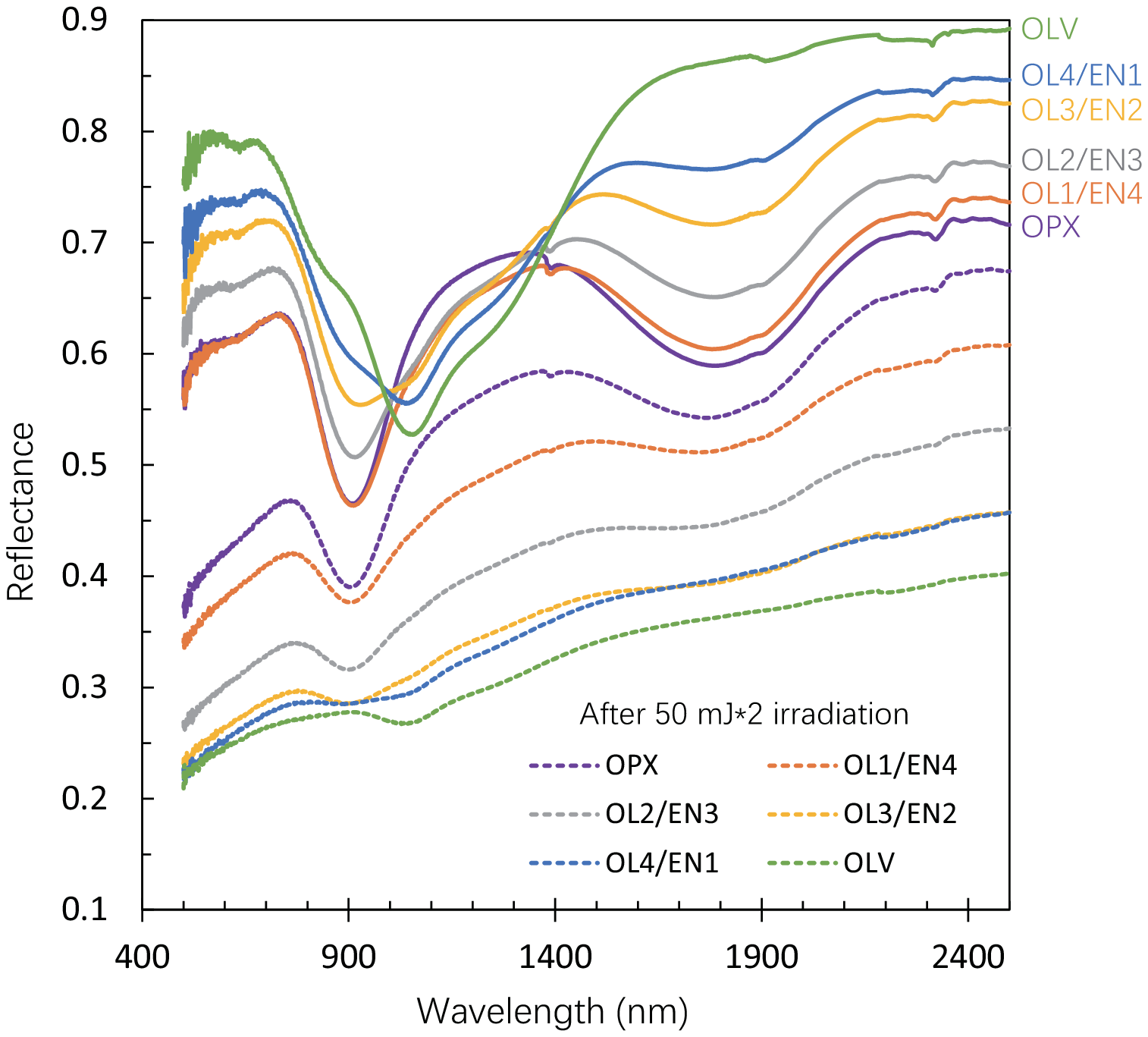}
		\caption{VNIR reflectance spectra of mixtures composed of olivine and orthopyroxene mixed with different proportions before (solid lines) and after (dashed lines) 50 mJ*2 laser irradiation. Narrow $\sim$1.41 and $\sim$2.33 $\mu$m absorption bands present in some spectra are due to minor alteration. The colors of spectral curves are the same as Figure \ref {Fig3}b.}
		\label{Fig4}
	\end{figure}

	A nanosecond pulsed laser (1 $\sim$ 50 mJ incident energy per pulse) is utilized as a micrometeorite bombardment simulation system with pulse frequency of 15 Hz. The laser spot is focused to a circle with a diameter of 0.5 mm using a lens and its moving speed is 1 mm s$^{-1}$. Researches show that after low-energy laser irradiation, the reflectance of olivine has a much greater change than that of pyroxene. Even after multiple irradiations, the depth of the pyroxene absorption band does not change much (\citealt{yamada1999simulation, jiang2019bi}). In order to better simulate the space weathering, we choose the maximum energy density (50 mJ) to irradiate the powder samples twice. The first irradiation is to translate in the vertical direction, and the second time is to translate in the horizontal direction. Each irradiated sample is measured immediately after irradiation. Original and irradiated reflectance spectra with the 2.14 $\mu$m artificial peak removed are shown in Figure \ref{Fig4}.

	The irradiation energy density is set to be 3000 mJ mm$^{-2}$ for each sample in our experiments. Due to the reflection and absorption of the glass, the energy transmission rate through the vacuum chamber glass window is about 90\%. Considering energy loss as the presence of the glass lens, glass slide and aluminum sample holder, as well as the barrier effect caused by the npFe$^{0}$ vapor deposition and the mineral particles adhered to the inner surface of the glass slide, we assume that 10\% of the energy is absorbed by the samples, so the energy deposition is about 300 mJ mm$^{-2}$. Total energy deposition rate by dust impacts at 1 AU is  estimated to be about 10$^{-3}$ J m$^{-2}$ yr$^{-1}$ based on the  calculations of \cite{yamada1999simulation} and \cite{sasaki2001production}. Possibly the irradiation of samples in our experiments corresponds to about 300 Myr in the space.

	%---------------------------Figures 4----------------------------
	%-----------------------------------------------------------------

	\subsection{The Modified Gaussian Model (MGM)}
	\label{subsec:The Modified Gaussian Model (MGM)}
	
	Originally developed by \cite{Sunshine1990Deconvolution}, MGM is a mathematical process based on the absorption of electronic transitions and is a statistical method for studying probability distribution. It is a mathematical procedure for deconvolving the superposed absorption features (\citealt{Sunshine1993Estimating}).The MGM program can be obtained from Brown University $\footnote{\url{http://www.planetary.brown.edu/mgm/}}$. So far, some complex minerals and rock samples have been explored with the MGM approach (e.g., \citealt{noble2006using, clenet2011new, pinet2016mgm}). However, we need more efforts to improve our spectral modeling and interpretation capability when dealing with the spectra of natural unknown mafic bodies.

\begin{figure}
		\centering
		\includegraphics[width=12.0cm, angle=0]{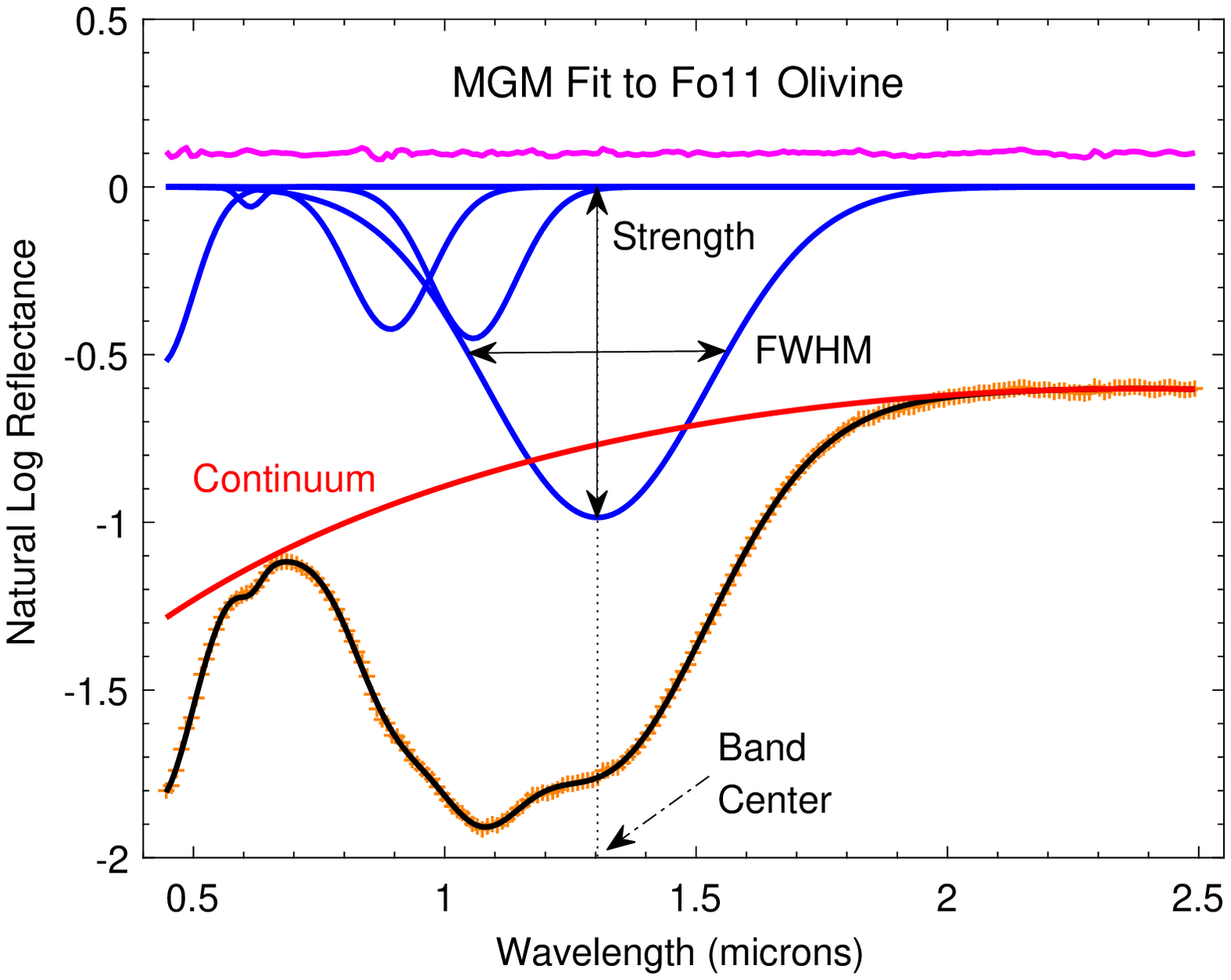}
		\caption{Four significant characteristic parameters for MGM fitting. For the sake of clarity, the residuals (pink line) are shifted by +0.1 on the ordinate.}
		\label{Fig5}
	\end{figure}

	To elaborate on this model, four significant characteristics are used for interpreting and exploiting the MGM, which are: band center, FWHM (full width at half maximum), band strength, and continuum, as illustrated in Figure \ref {Fig5}. In MGM program, natural logarithm of each spectral curve is modeled in energy domain (wave number), which means in wavelength domain a natural logarithm spectrum can be deconvolved into multiple Gaussian distributions with a continuum. MGM is modeled as:

	%------------------------Eq. 1---------------------------------
	\begin{equation}
		\label{Eqn1}
		\ln\left [ R\left ( \lambda  \right ) \right ]=C\left ( \lambda \right )+ \sum_{i=1}^{N}S_{i}\cdot \exp\left [ -\frac{\left ( \lambda ^{-1}-\mu _{i}^{-1} \right )^{2}}{2\sigma _{i}^{2}} \right ]
	\end{equation}
	
	where $\lambda$ is the wavelength, \emph{R} ($\lambda$) is the reflectance at wavelength $\lambda$, \emph{S} is the band strength, $\mu$ is the band center, $\sigma$ is the band width, and \emph{N} is the number of bands. \emph{C}(\emph{$\lambda$}) is the function of the continuum in wavelength space.
	%---------------------------------------------------------------------

	Here the continua are fitted by using the logarithm of second-order polynomial. Continua are modeled close to the inflection regions of each absorption band, and we set an uncertainty within 20\% for each parameter:
	
	%------------------------Eq. 2---------------------------------
	\begin{equation}
		\label{Eqn2}
		C\left ( \lambda  \right )=\ln \left ( c_{0} +c_{1} \lambda+c_{2} \lambda^{2}\right )
	\end{equation}
	
	\begin{center}
		$\lambda$= wavelength,  $c_{0}, c_{1}, c_{2}$= constants
	\end{center}
	%---------------------------------------------------------------

	%---------------------------Figures 5----------------------------
	%-----------------------------------------------------------------

	\section{Results and Discussions}
	\label{sect:Results and Discussions}
	
	\subsection{MGM Parameters of Olivine and Orthopyroxene}
	\label{subsec:MGM Parameters of Olivine and Orthopyroxene}
	
\begin{figure}
		\centering
		\includegraphics[width=15.0cm, angle=0]{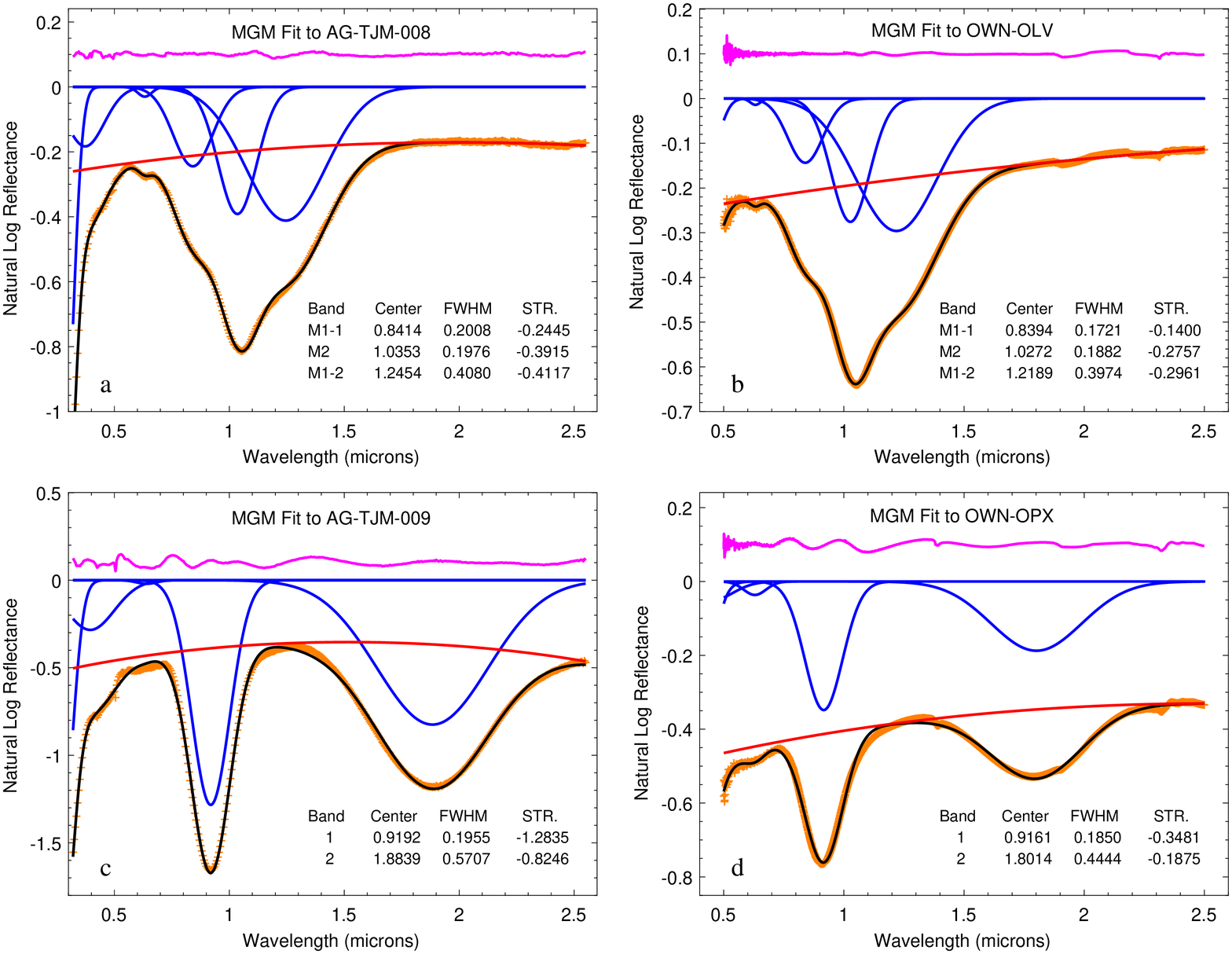}
		\caption{The comparison of MGM fitting results for the spectral curves with olivine and orthopyroxene obtained from RELAB database (panel (a): RELAB olivine and panel (c): RELAB orthopyroxene) and our experiments (panel (b): OWN olivine and panel (d): OWN orthopyroxene).}
		\label{Fig6}
	\end{figure}

	In VNIR regions, olivine spectra show a broad and complex absorption feature near 1.0 $\mu$m. The Gaussian curves presented in the top panels of Figure \ref {Fig6} indicate that olivine spectra are composed of three individual absorption bands, they are around 0.84 $\mu$m, 1.03 $\mu$m and 1.23 $\mu$m respectively, which are assigned to the M1–1, M2 and M1–2 sites in olivine crystal (\citealt{burns1970crystal}). The width of each Gaussian absorption band in olivine spectra is related to the iron abundance in olivine crystal and their ratio can be applied to constrain the parameters in MGM (\citealt{Sunshine1998Determining}). Preliminary analyses of band strength derived by the MGM from olivine reflectance spectra show the bandwidths of M1-1 and M2 are similar but the width of M1-2 is about twice as large as M1-1 and M2. Analogously, the absorption strength can also be evaluated as a function of composition. Band strength of M2 and M1-2 are similar and  about twice of M1-1.

	VNIR reflectance spectra of low-Fe orthopyroxene are dominated by the absorptions near 0.92 $\mu$m and 1.85 $\mu$m (bottom panels in Figure \ref {Fig6} ). Each absorption is primarily caused by one electronic transition absorption band (\citealt{Sunshine1990Deconvolution}). The strength of $\sim$0.92 $\mu$m band is stronger than that of $\sim$1.85 $\mu$m band, whereas the band width of $\sim$1.85 $\mu$m absorption is more than twice as large as $\sim$0.92 $\mu$m absorption band.

	%---------------------------Figures 6----------------------------
	%-----------------------------------------------------------------

	\subsection{MGM Parameters of Olivine-Orthopyroxene Mixtures}
	\label{subsec:MGM Parameters of Olivine-Orthopyroxene Mixtures}
	
\begin{figure}
		\centering
		\includegraphics[width=15.0cm, angle=0]{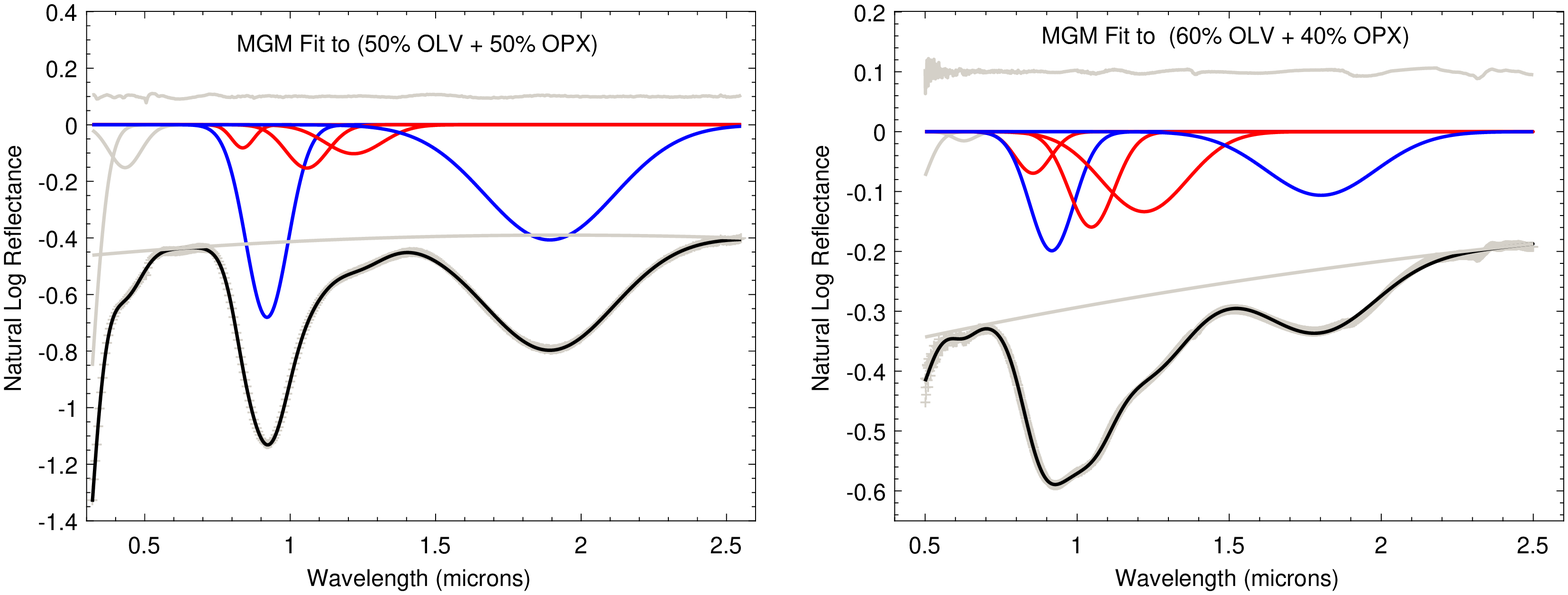}
		\caption{MGM fitting results for the samples composed of olivine and orthopyroxene mixed with different proportions. The fitting curves with the OPX absorption features are plotted as the blue curve. The red curve represents the olivine absorption bands fitting by using the Gaussian function.}
		
		\label{Fig7}
	\end{figure}

	The initial MGM parameters of the olivine and orthopyroxene in the two end-member mixture can be used as constraints for further analysis of mixture spectra. We utilize these parameters to initialize the MGM and an intermediate ratio mixture in each group for spectral deconvolution. They are sample IDs: AG-TJM-019 (50\% OLV + 50\% OPX) in RELAB and OWN-OL3/EN2 (60\% OLV + 40\% OPX) in our experiment. The starting and final parameters of these two spectra in MGM program are listed in Table \ref {Tab3} and the fitting results are shown in Figure \ref {Fig7}.

	%------------------------Table   3---------------------------------
	\begin{table}
		\bc
		\begin{minipage}[]{150mm}
			\caption[]{MGM modelling parameters of mixture samples spectra
				\label{Tab3}}\end{minipage}
		\setlength{\tabcolsep}{2.5mm}
		\small
		\begin{tabular}{ccc|ccc|c}
			\hline\noalign{\smallskip}
			\multicolumn{2}{c}{Sample ID} &\multicolumn{2}{c}{AG-TJM-019}& &\multicolumn{2}{c}{OWN-OL3/EN2}\\
			\hline\noalign{\smallskip}
			\multicolumn{2}{c}{Abundance} &\multicolumn{2}{c}{50\% OLV + 50\% OPX}& &\multicolumn{2}{c}{60\% OLV + 40\% OPX}\\
			\hline\noalign{\smallskip}
			\multicolumn{2}{c}{\multirow{2}*{Gaussian Fitting Parameters} }&Starting	&Final	& &	Starting&Final\\
			&	&Parameters	& Parameters& &Parameters& Parameters \\
			\hline\noalign{\smallskip}
			&Parameter 1 	&6.00E-01&	6.17E-01&&		6.60E-01&	6.57E-01\\
			Continua   &Parameter 2	&8.50E-05&	8.49E-05&&		1.10E-05&	1.10E-05\\
			&Parameter 3	&-2.30E-08&	-2.50E-08&&		-1.50E-08&	-1.68E-08\\
			&	&	&	&	&	&	\\
			&Center	&	840&	836.8&&		840&	859.1\\
			M1-1	&FWHM	&	150&	102.9&&		150&	141.7\\
			&Strength&	-0.08&	-0.802&&	-0.07&	-0.069\\
			&	&	&	&	&	&	\\
			&Center	&	1035&	1056.1&&		1030&	1050.0\\
			M2		&FWHM	&	150&	183.6&&		170&	168.6\\
			&Strength&	-0.15&	-0.152&&	-0.13&	-0.141\\
			&	&	&	&	&	&	\\	
			&Center	&	1245&	1221.6&&		1220&	1211.4\\
			M1-2	&FWHM	&	300&	258.3&&		320&	353.2\\
			&Strength&	-0.16&	-0.117&&		-0.15&	-0.145\\
			&	&	&	&	&	&	\\
			&Center	&	920&	919.6&&		920&	919.1\\
			Band 1	&FWHM	&	200&	176.8&&		200&	179.5\\
			&Strength&	-0.65&	-0.686&&		-0.20&	-0.203\\
			&	&	&	&	&	&		\\
			&Center	&	1885&	1888.5&&		1800&	1800.1\\
			Band 2	&FWHM	&	500&	541.2&&		450&	439.5\\
			&Strength&	-0.40&	-0.422&&		-0.12&	-0.114\\
			\hline\noalign{\smallskip}
			\multicolumn{2}{c}{Root Mean Square Error} &\multicolumn{2}{c}{2.523E-03}& &\multicolumn{2}{c}{3.588E-03}\\
			\hline\noalign{\smallskip}
		\end{tabular}
		\ec
		\tablecomments{0.97\textwidth}{Band center and FWHM are in the unit of nanometers. The final parameters are rounded off to proper precision. Band 1 is for absorption near 1.0 $\mu$m in OPX, while Band 2 is for absorption near 2.0 $\mu$m in OPX.}
	\end{table}
	%----------------------------------------------------------------

	The results show that when olivine and orthopyroxene are mixed in equal amount, orthopyroxene dominates the spectral absorption features of the mixture. As shown in Figure \ref {Fig6} and Figure \ref {Fig7}, the center positions of the orthopyroxene's absorption bands are relatively stable, the strength is close to half of the pure orthopyroxene and the ratio of FWHM parameters remains the same. While, parameters of the three Gaussian absorption bands in olivine spectra change a lot, the strength of each absorption band is about 1/3 of the pure olivine mineral, and the strongest absorption M1-2 band decrease sharply. When the content of olivine is more than orthopyroxene, the spectral curve of the mixture near 1.0 $\mu$m region show a broad asymmetric complex absorption feature, and the absorption features of olivine dominate the spectra curve between 1.0 to 1.5 $\mu$m regions.

	%---------------------------Figures 7----------------------------
	%-----------------------------------------------------------------

	\subsection{Modified BAR of Olivine-Orthoyroxene Mixtures}
	\label{subsec: Modified BAR  of Olivine-Orthoyroxene Mixtures}
	
\begin{figure}
		\centering
		\includegraphics[width=12.0cm, angle=0]{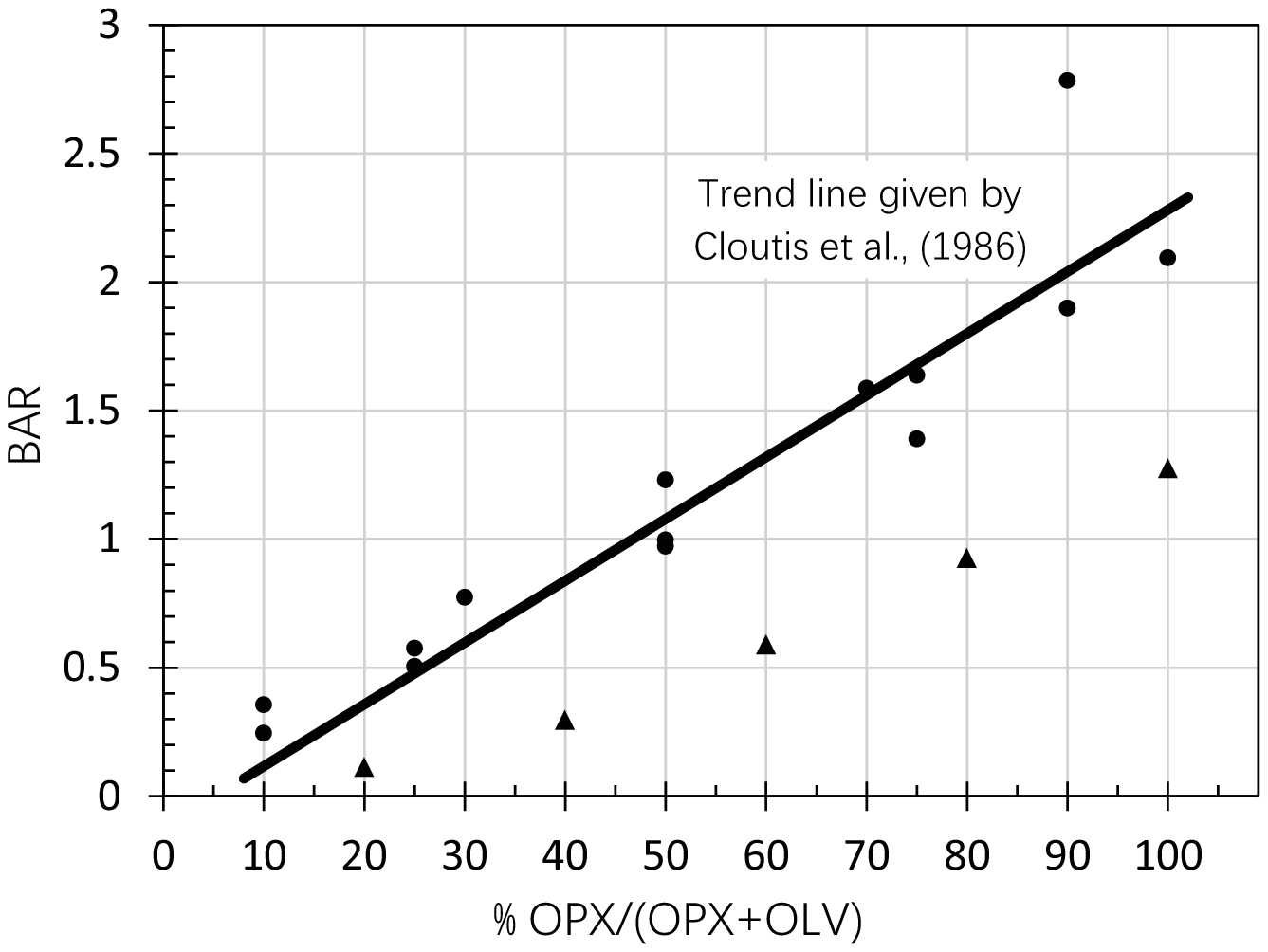}
		\caption{BAR values derived from the spectra with olivine–orthopyroxene mixtures utilized in this study. The solid line represents the trend line given by \cite{cloutis1986calibrations}. Black dots and triangles denote the results estimated from RELAB and our original experiment samples respectively.}
		\label{Fig8}
	\end{figure}

	We also utilize the band area ratio (BAR) method (\citealt{cloutis1986calibrations}) to analyze these spectra. BAR method does not need MGM processing and it is generally considered as a very useful indicator for interpreting the olivine–orthopyroxene mixture spectra, which is nearly independent of mineral composition and mineral grain size (\citealt{cloutis1986calibrations}). However, we found that the BAR values of each mixture sample group show a well linear trend, but not all of them follow the trend line given by \cite{cloutis1986calibrations}. In the calculation results of some samples, especially for the BAR values of our newly acquired spectral curves, the deviation between the estimated mineral abundance value and its true value maybe more than 30\% (Figure \ref {Fig8}).

	The deviations are mainly caused by using different spectrometers as stated in \cite{hapke2012theory}, band depths and band shapes depend on illumination and viewing geometry. It can cause differences in band shapes between spectra of the same material measured bidirectionally and using an integrating sphere to measure the directional–hemispherical reflectance. Therefore, we should use this trend line with caution, especially when the spectrometers and/or the viewing geometry used for measurement are different and the abundances and/or end-member chemistries are unknown. Fortunately, the well linear trend shown by each BAR parameter group provides a direction for solving this problem. As shown in Figure \ref {Fig6} and Figure \ref {Fig7}, M1-2 band ($\sim$1.23 $\mu$m) in olivine absorption features is able to be distinguished from the absorption band of orthopyroxene. The absorption strength of olivine M1-2 band and overall absorption feature near 1.0 $\mu$m contain the information about the relative content of olivine and orthopyroxene. It indicates that we can use the MGM fitting results to estimate the mineral abundance.

	%---------------------------Figures 8----------------------------
	%-----------------------------------------------------------------

\begin{figure}
		\centering
		\includegraphics[width=11.0cm, angle=0]{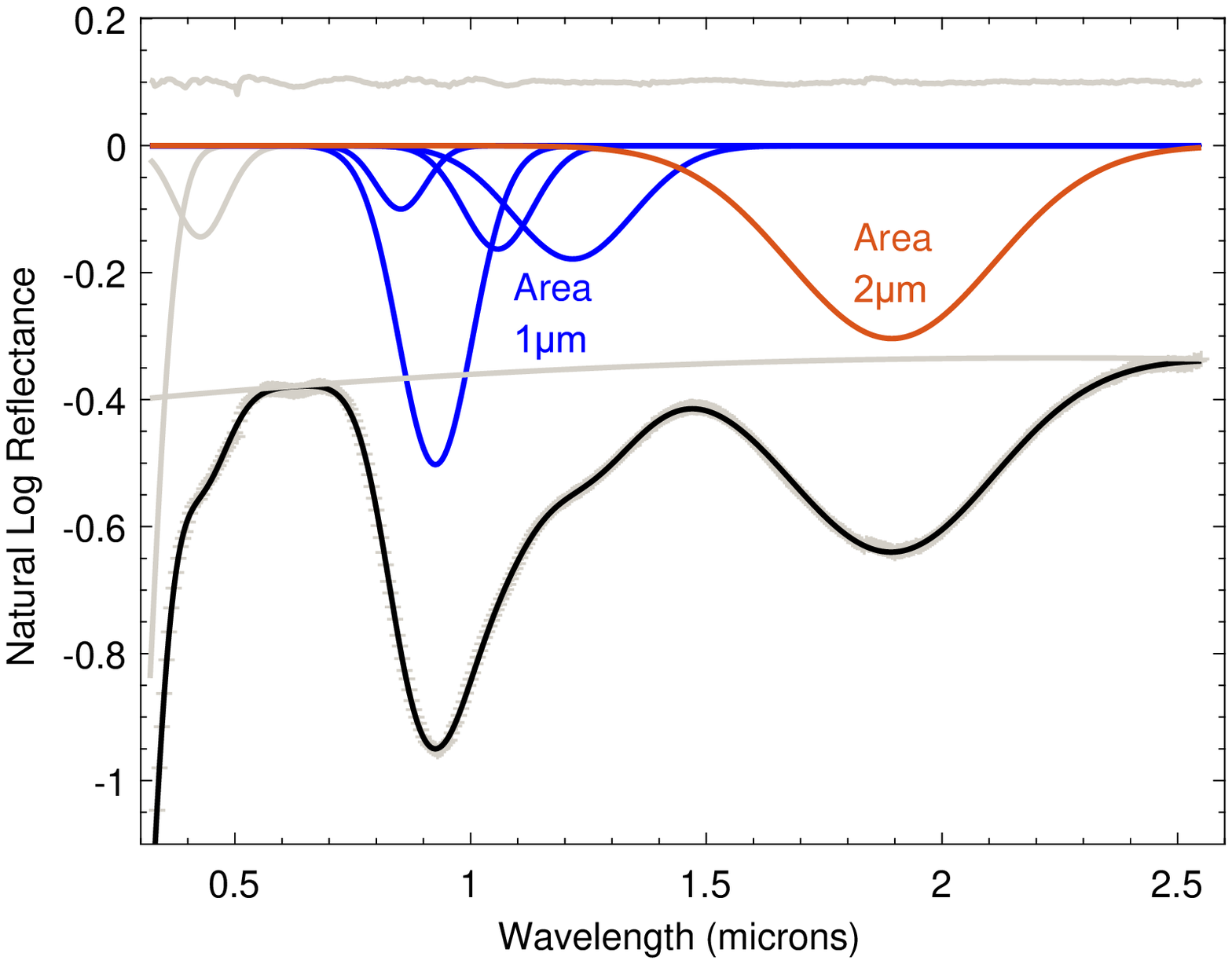}
		\caption{The spectral diagnostic feature of modified band area ratio (MBAR) derived from MGM fittings. For the obvious absorption feature close to 1.0 $\mu$m, the area of this band composed of four Gaussian curves (shown as blue) is defined as Area\_1$\mu$m. For the band at around 2.0 $\mu$m, this feature is made by single Gaussian curve with orthopyroxene (orange line) and then defined as Area\_2$\mu$m.}
		\label{Fig9}
	\end{figure}

	Firstly, we utilize MGM to obtain the absorption features of olivine and orthopyroxene in the mixtures. Secondly, a modified band area ratio (MBAR) indicator is obtained by using MGM fitting result parameters. The MBAR is modeled as:
	
	%------------------------Eq. 3---------------------------------
	\begin{equation}
		\label{Eqn3}
		MBAR = Area\_2\mu m/Area\_1\mu m
	\end{equation}
	
	%---------------------------------------------------------------
	
	where, \emph{Area} is calculated with an approximation and it is defined as:
	
	%------------------------Eq. 4---------------------------------
	\begin{equation}
		\label{Eqn4}
		Area=\sum_{i=1}^{N} 1.065* FWHM_i*STR_i 
	\end{equation}
	
	%---------------------------------------------------------------
	
	where,	1.065 is the factor to get the integral of Gaussian curve by using \emph{FWHM}. \emph{STR} is the absolute value of absorption strength. There are 4 absorption bands in Area\_1$\mu$m, which are the three diagnostic absorption features of olivine and one absorption of orthopyroxene (that is Band 1). There is only one orthopyroxene absorption band (Band 2) in Area\_2$\mu$m. The illustration in Figure \ref {Fig9} explains these definitions in detail.

	%---------------------------Figures 9----------------------------
	%-----------------------------------------------------------------

\begin{figure}
		\centering
		\includegraphics[width=12.0cm, angle=0]{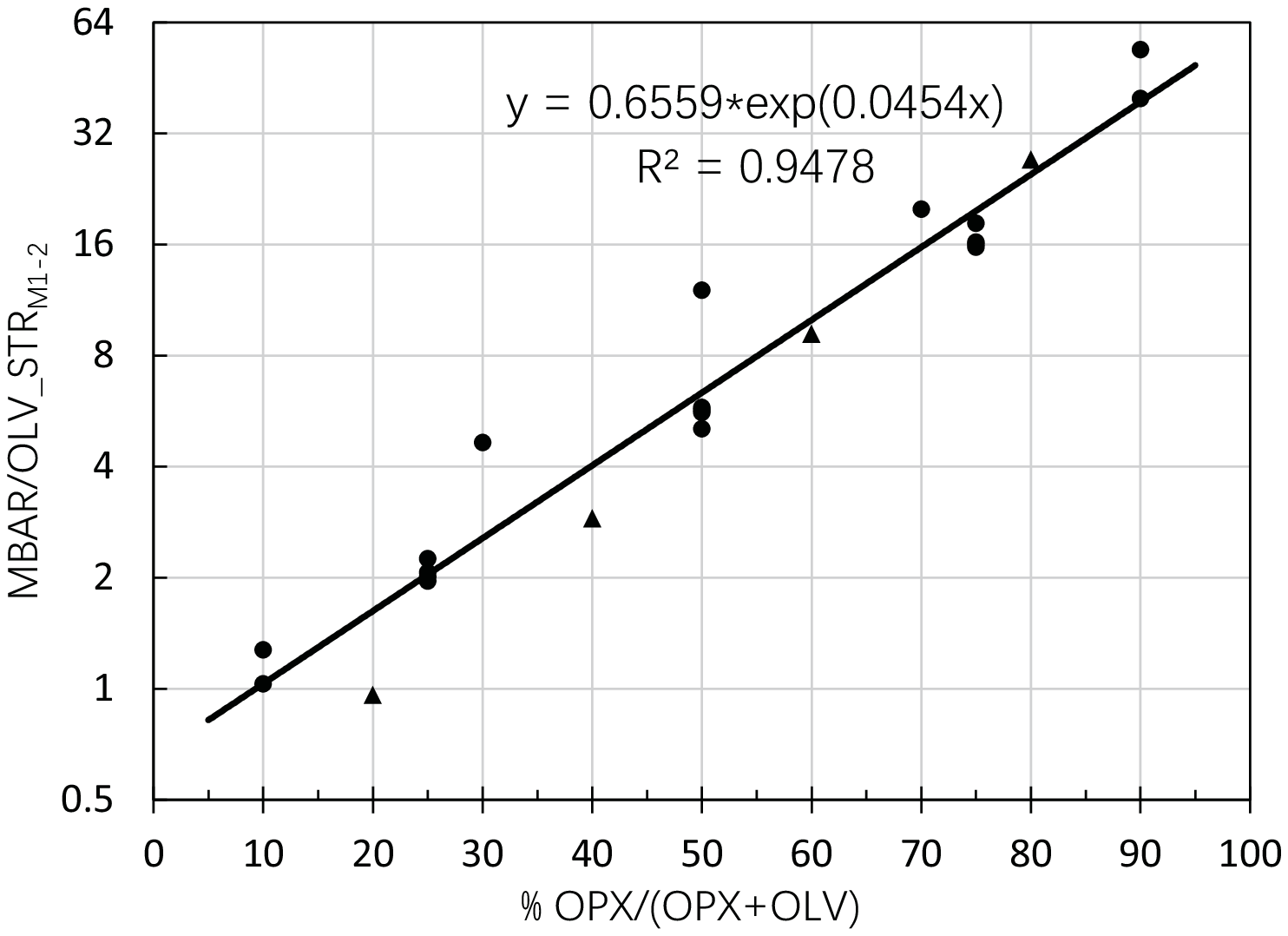}
		\caption{A plot of MBAR/OLV\_STR$_{{\rm M}1-2}$ ratio versus orthopyroxene abundance. The solid line represents a linear least-square fit for all data with the orthopyroxene abundance between 10\% and 90\%. The notations of data points are the same as Figure \ref {Fig8}.}
		\label{Fig10}
	\end{figure}

	Finally, MBAR can be regarded as the absorption intensity coefficient of the two main absorption bands in the olivine–orthopyroxene mixture spectra. As is mentioned above, M1-2 band in olivine spectra are easy to extract from the mixture spectral curves. Dividing MBAR by the olivine M1-2 band absorption strength value (abbreviated as: MBAR/OLV\_STR$_{{\rm M}1-2}$) can get the end-member composition information. The result values show a good pattern on the estimation results, as the orthopyroxene abundance increases, the MBAR value increases, the OLV\_STR$_{{\rm M}1-2}$ value decreases and the ratio data of the two values show an exponential growth trend. In order to show their changing patterns clearly, we use an exponential ordinate to display them. Figure \ref {Fig10} shows the calculation results and the regression equation. The overall error in predicting the mixture composition based on MGM fitting results is less than 15\%.

	%---------------------------Figures 10----------------------------
	%-----------------------------------------------------------------

\begin{figure}
		\centering
		\includegraphics[width=15.0cm, angle=0]{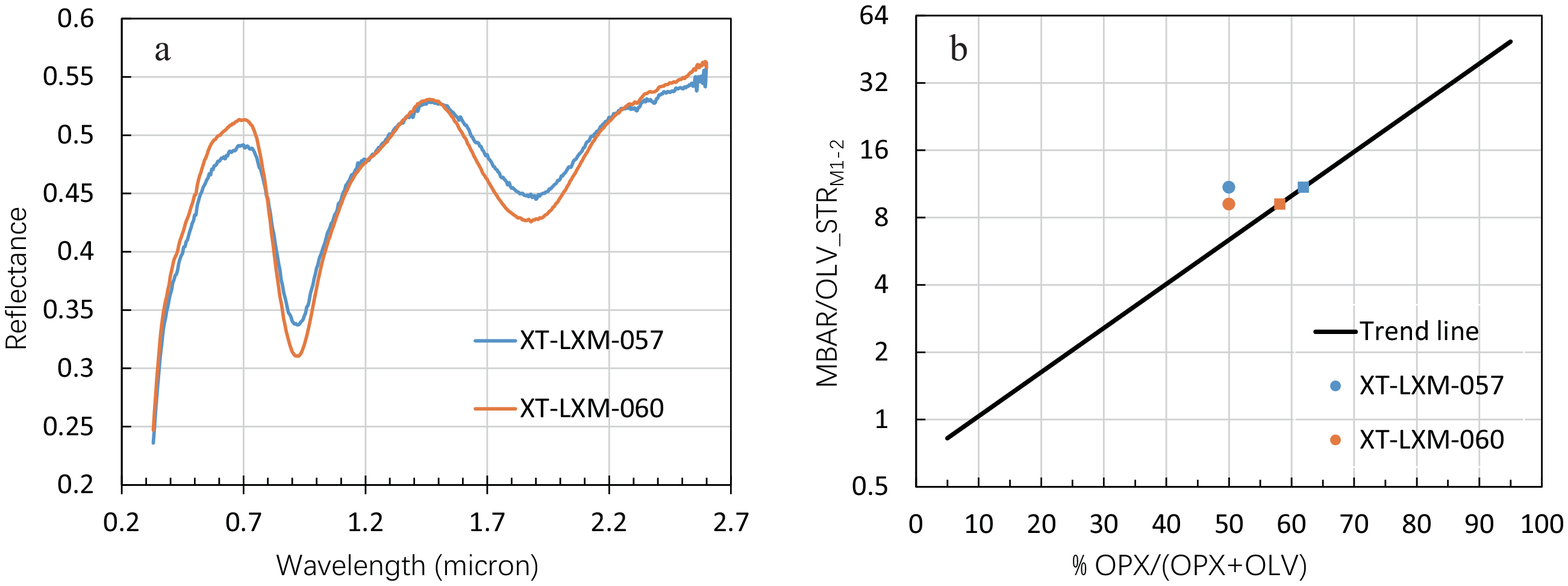}
		\caption{Spectral reflectance of two RELAB olivine–orthopyroxene mixture samples and mineral abundance estimation results obtained by MGM. Dot represents the true value and square represents the estimated value.}
		\label{Fig11}
	\end{figure}

	Here, to test the deconvolution analysis and the trend lines described above, we choose two olivine-orthopyroxene mixture samples with the same end-member minerals as examples (RELAB ID: XT-LXM-057 and XT-LXM-060, whose grain sizes are 32 $\sim$ 63 $\mu$m, with 50\% olivine + 50\% orthopyroxene). These two reflectance spectra are measured by placing identical samples in different sample dishes. XT-LXM-057 is placed in an aluminum dish coated with black Teflon and XT-LXM-060 is placed in a Nicolet dish. Two different spectral curves are obtained by changing the sample holding dishes. Their reflectance spectra are shown in Figure \ref {Fig11}a. The deconvolution fitting results are obtained by using the MGM and the mineral abundance estimation results are shown in Figure \ref {Fig11}b. MBAR of the two mixture samples are 0.850 and 0.861, olivine M1-2 band absorption intensity values are $\sim$0.078 and $\sim$0.094. MBAR/OLV\_STR$_{{\rm M}1-2}$ values are 10.9 and 9.2 respectively. The orthopyroxene abundance estimation results are 61.9\% and 58.1\%. Compared to the true value (50\%), these two results are all within the accuracy range of 15\% given above.

	%---------------------------Figures 11----------------------------
	%-----------------------------------------------------------------

	\subsection{Fitting Results of Irradiated Samples}
	\label{subsec:FRoIS}
	
\begin{figure}
		\centering
		\includegraphics[width=15.0cm, angle=0]{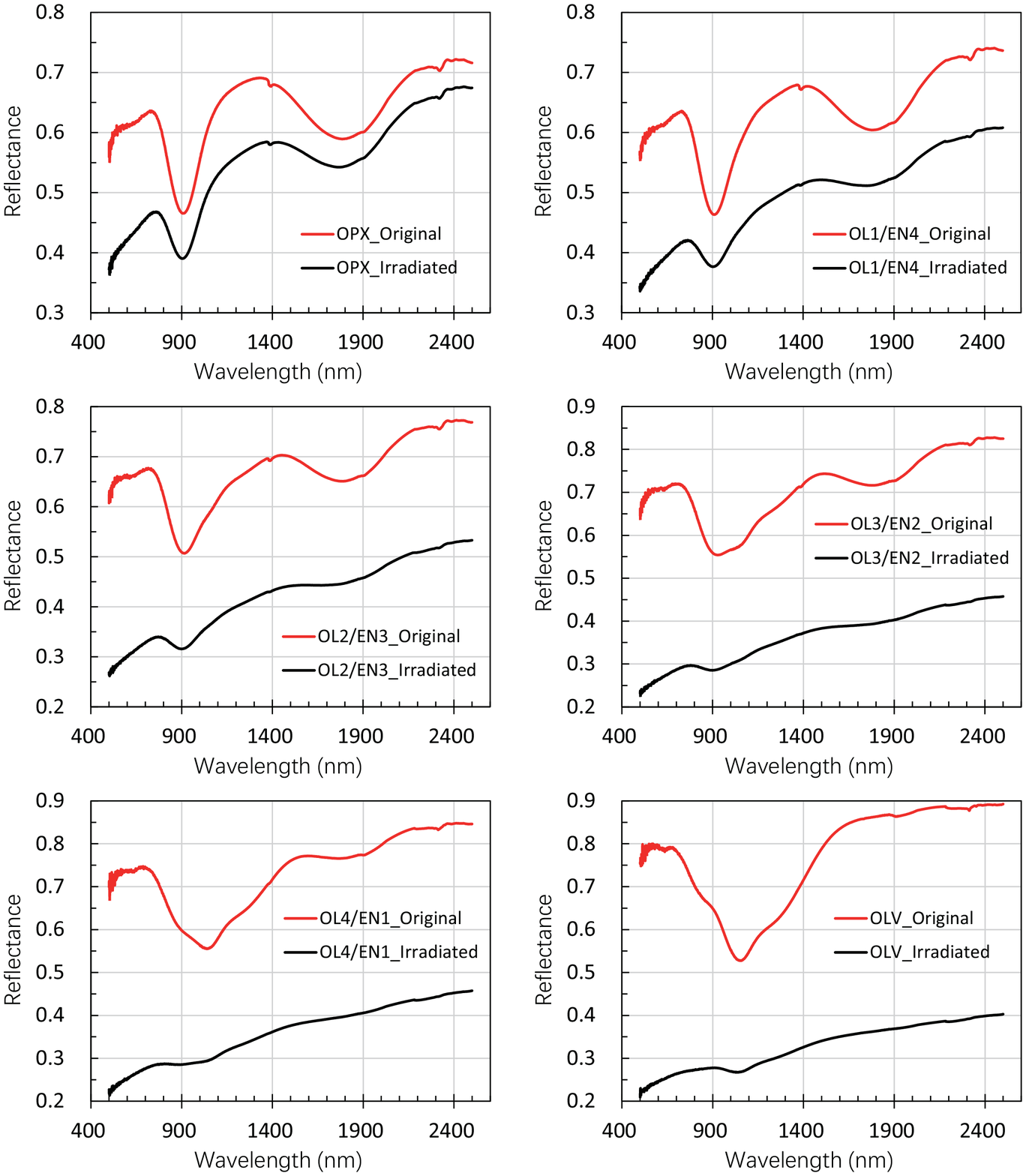}
		\caption{Comparison of the spectral curves composed of olivine-orthopyroxene admixtures mixed with different proportions and their irradiated results.}
		\label{Fig12}
	\end{figure}

	Some RELAB samples irradiated with the pulsed laser to simulate the space weathering effect are mentioned in Table \ref {Tab2}, but it seems no significant spectral alteration with their spectral curves. However, for our samples, the fresh spectra and irradiation spectra by high energy laser show obviously difference (Figure \ref {Fig12}), which are suitable for further study. After irradiation, the spectral features turn weaker and shallower. Absorption intensity and albedo of the visible region decrease faster than the NIR region, whose behaviors redden and darken the irradiated VNIR spectra. In a high-intensity or long-term space weathering environment, the spectra of pyroxene appear more reddening, and olivine spectra appear more darkening. This is consistent with the results presented in \cite{yamada1999simulation} and \cite{jiang2019bi}.

	%---------------------------Figures 12----------------------------
	%-----------------------------------------------------------------

\begin{figure}
		\centering
		\includegraphics[width=15.0cm, angle=0]{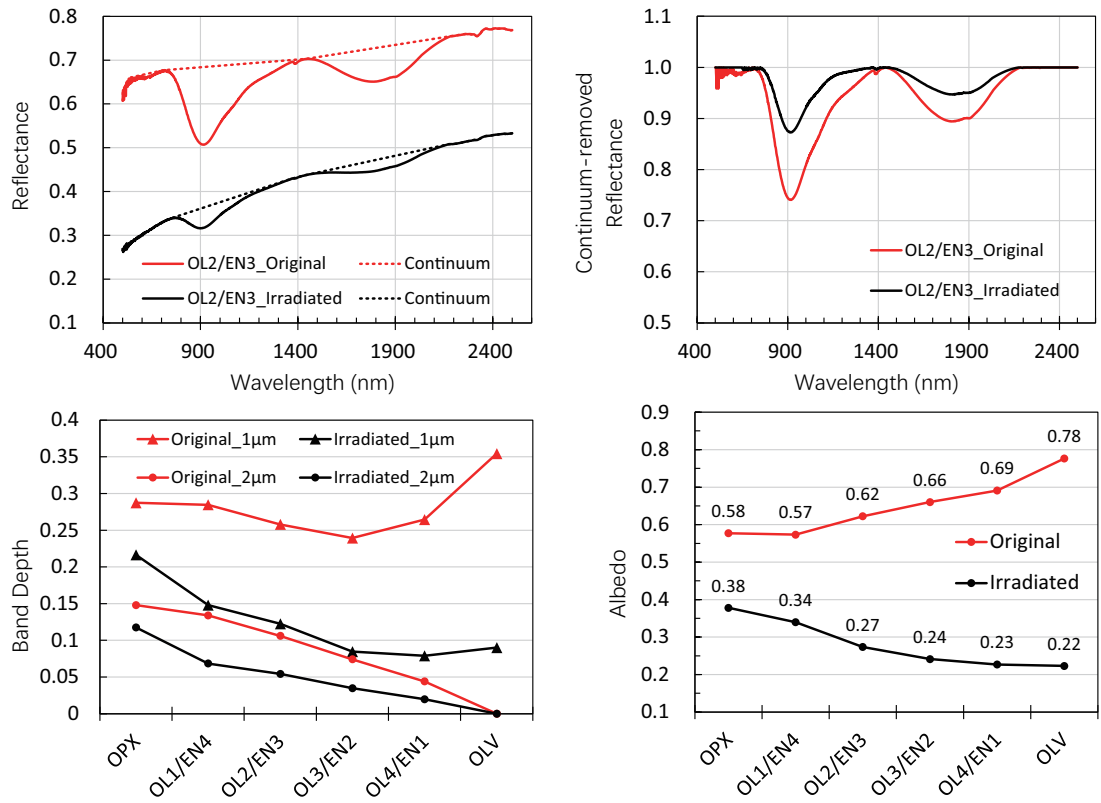}
		\caption{Variations of band-depths and albedos (average of the spectral reflectance ranging from 0.5 to 0.8 $\mu$m) of olivine-orthopyroxene mixtures before and after pulsed-laser irradiation.}
		\label{Fig13}
	\end{figure}

	A widely utilized continuum remove method is employed to compare the spectral absorption features and the band depth of these samples before and after pulsed laser radiation (\citealt{clark1984reflectance}). Band center positions of the continuum-removed spectra before and after laser irradiation are compared in Table \ref {Tab4}, it is generally believed that continuum-removed band positions are less affected (shift a few nanometers) by space weathering (\citealt{pieters2000space, fu2012effects}). Figure \ref {Fig13} shows the parameter comparison of these original and irradiated spectral curves. The absorption features near 1.0 $\mu$m is the superposition of olivine and orthopyroxene. Obviously, the higher the proportion of olivine is, the more band depth decreases after laser irradiation. While the absorption features near 2.0 $\mu$m is only due to the spectral absorption of orthopyroxene, the spectral curves in this region change uniformly. A sub-figure shows the albedo value of these samples and here it uses the average of the spectral reflectance from 0.5 to 0.8 $\mu$m (close to visible region). This figure shows that orthopyroxene has stronger resistance to space weathering than olivine. On the whole, the higher content of olivine is in mixture, the weaker absorption features are in its irradiated spectra, which may make these spectra more difficult to identify and interpret.

	%------------------------Table   4---------------------------------
	\begin{table}
		\bc
		\begin{minipage}[]{150mm}
			\caption[]{Comparisons of the absorption band centers before and after irradiations
				\label{Tab4}}\end{minipage}
		\setlength{\tabcolsep}{2.5mm}
		\small
		\begin{tabular}{ccccccc}
			\hline\noalign{\smallskip}
			\multirow{2}*{Sample ID} &&\multicolumn{2}{c}{Original}& &\multicolumn{2}{c}{Irradiated}\\
			 			&&(1$\mu$m)&(2$\mu$m)&&(1$\mu$m)&(2$\mu$m)\\
			\hline\noalign{\smallskip}
			OWN-OPX     &&	912.16&	1795.26&&	913.45&	1803.38\\
			OWN-OL1/EN4 &&	911.52&	1801.50&&	912.16&	1805.26\\
			OWN-OL2/EN3 &&	915.38&	1804.01&&	915.38&	1809.04\\
			OWN-OL3/EN2 &&	931.16&	1824.32&&	922.55&	1808.41\\
			OWN-OL4/EN1 &&	1037.15& 1824.32&&	1041.11&1824.32\\
			OWN-OLV		&&	1056.17&	- - &&	1057.03& - -   \\
			\hline\noalign{\smallskip}
		\end{tabular}
		\ec
		\tablecomments{0.7\textwidth}{Band centers are in nm, values listed in this table are approximations.}
	\end{table}
	%----------------------------------------------------------------

	%---------------------------Figures 13----------------------------
	%-----------------------------------------------------------------

	%---------------------------Figures 14----------------------------
	\begin{figure}
		\centering
		\includegraphics[width=15.0cm, angle=0]{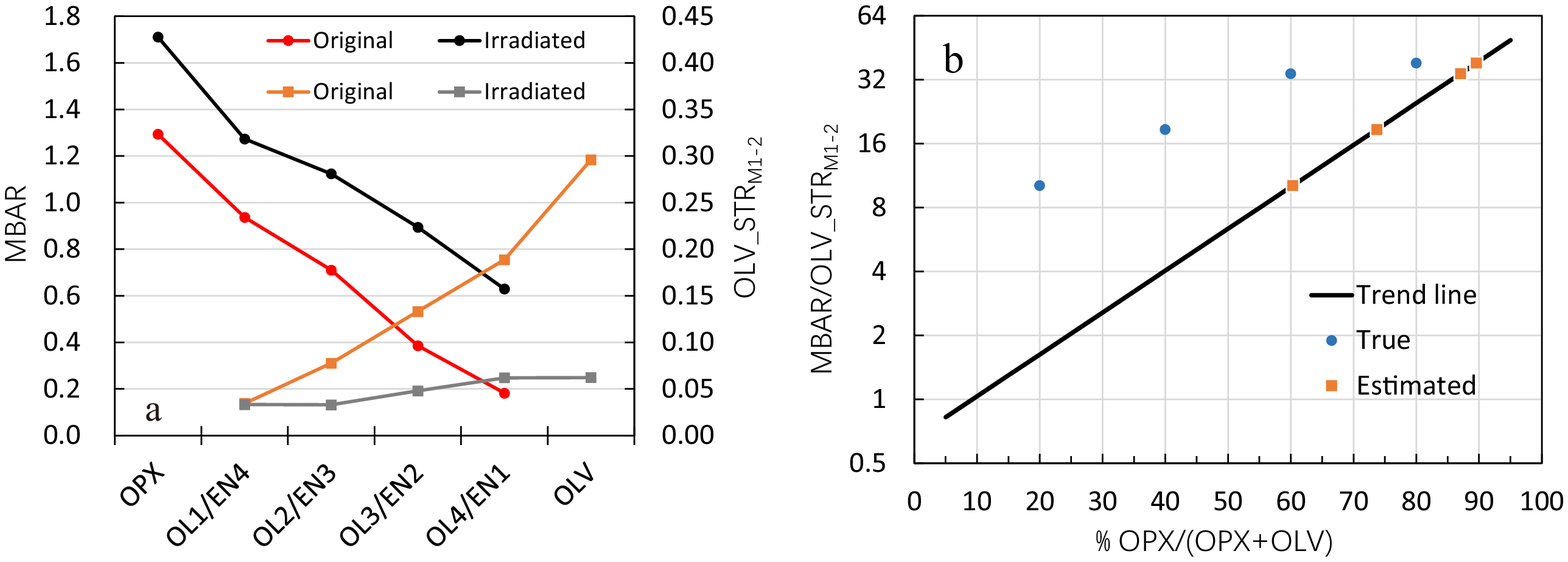}
		\caption{Deconvolution results of the olivine-orthopyroxene mixtures derived from MGM fittings. (a). The evolution of MBAR ratios (Band M1-2 absorption strengths) of olivine-orthopyroxene mixtures with different proportions before and after pulsed-laser irradiation. (b). Correlation of truthful and estimation abundances of the mixtures with pulsed-laser irradiation, the blue circles and orange squares represent the truthful and estimation abundances of these mixtures respectively.}
		\label{Fig14}
	\end{figure}
	%-----------------------------------------------------------------

	Also, we deconvolve these space weathered spectra based on the prior mineral knowledge and the constraints mentioned above. After pulsed laser irradiation, the MBAR of these samples all increase (Figure \ref {Fig14}a). It can be inferred that the spectra of the short wavelength are more susceptible to space weathering than the longer wavelength in the VNIR region. What is more, as olivine is more susceptible to space weathering, the spectral absorption strength of olivine will reduce significantly. The more olivine content is in the mixtures, the greater ratio decrease of Band M1-2 absorption strength (Figure \ref {Fig14}a), which will further increase the deviation in the mixture composition prediction results. MGM deconvolution results of these irradiated mixtures are shown in Figure \ref {Fig14}b. The distance between two points in the horizontal direction represents the deviation value of the prediction result. With the increase of olivine (decrease of orthopyroxene) content, the deviation of estimate results also increases.

	In addition, an extended calculation shows that it will get a slightly larger MBAR/OLV\_STR$_{{\rm M}1-2}$ value than the original spectrum when using MGM to deconvolve the continuum-removed spectrum, that means it will slightly underestimate the olivine abundance. This finding will be used later. All these results remind us that we cannot be too cautious when identifying and interpreting the spectra of space weathered celestial bodies. If the effect of space weathering is not taken into account, MGM deconvolution results will usually underestimate the proportion of olivine in the mixtures.

	\subsection{Estimating the Composition of Olivine-Orthopyroxene Asteroids}
	\label{subsec:Estimating the Composition of Olivine-Orthopyroxene Asteroids}
	
	The research results obtained above are very promising and they can be used to study the space weathered celestial bodies, such as asteroids or lunar rocks. Asteroids are considered to be smaller and rotating faster than solid planets, with minimal gravity, these asteroids have thinner regolith (\citealt{chapman2004space}). Q-type and S-type asteroids are characterized by spectra with moderate silicate absorption features near 1.0 and 2.0 $\mu$m (\citealt{chapman2004space, demeo2015compositional}). \cite{vernazza2008compositional} and \cite{thomas2014physical} analyzed the VNIR spectra of many Q-type and S-type asteroids and reported that most of them have spectral properties similar to LL chondrites, whose main minerals are hypersthene and olivine. Here we choose some asteroids dominated by olivine-orthopyroxene for further study to estimate their mineral compositions.

	\subsubsection{Spectral Deconvolution on Olivine-Orthopyroxene Asteroids}
	\label{subsubsec:Olivine-Orthopyroxene Asteroids}
	
	The center positions of 2.0 $\mu$m absorption band are useful for determining the compositions of pyroxene. Low-Fe orthopyroxene spectra show that the absorption band center near 2.0 $\mu$m is ranging from 1.80 to 1.95 $\mu$m (\citealt{klima2007spectroscopy}). These informations can help us to search the suitable asteroid spectra for studying their mineralogical characterization. Six asteroids with high quality spectra are selected, they are: (4) Vesta, (15) Eunomia, (158) Koronis, (631) Philippina, (2335) James and (4954) Eric. The VNIR spectra of these asteroids are obtained from the Small Main-Belt Asteroid Spectroscopic Survey (SMASS), which are presented on the MIT planetary spectroscopy website $\footnote{\url{http://smass.mit.edu/home.html}}$.

\begin{figure}
		\centering
		\includegraphics[width=15.0cm, angle=0]{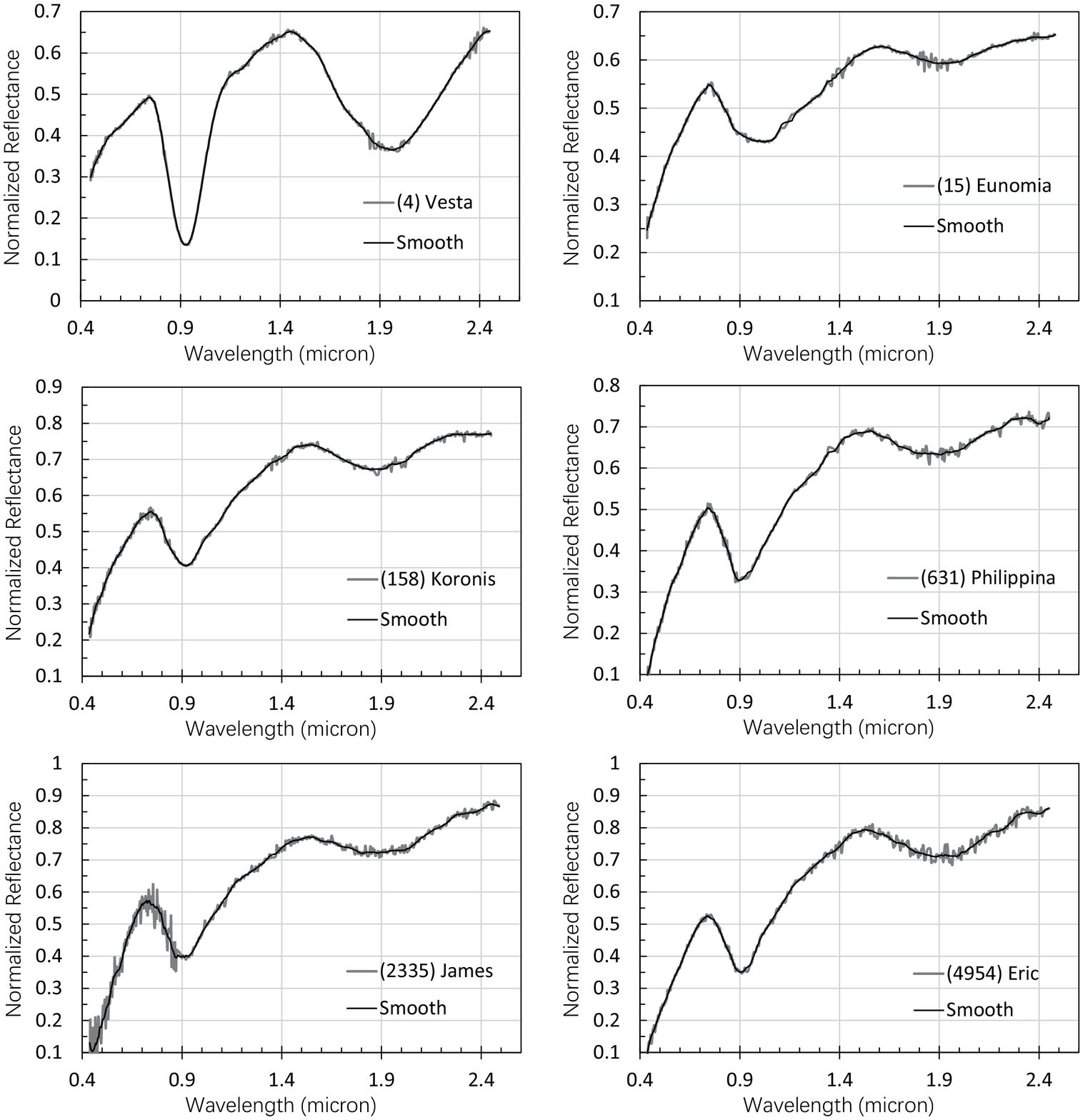}
		\caption{VNIR spectra of six asteroids (4) Vesta, (15) Eunomia, (158) Koronis, (631) Philippina, (2335) James and (4954) Eric. These spectra are normalized at 0.55 $\mu$m and shifted downward to suitable positions.}
		\label{Fig15}
	\end{figure}

	Absolute spectral albedos are difficult to obtain with accuracy in many remote sensing situations, so all spectra from MIT are normalized to unity at 0.55 $\mu$m. This work smoothed and shifted these asteroids spectra reflectance curves to acclimatize to the MGM. First, the spectra are smoothed using the Savitzky-Golay (SG) filter procedure (order = 3 and framelen = 25) (\citealt{Savitzky1964Smoothing}). Then, the spectral curves should be shifted to the suitable positions: if the maximum value of the asteroid albedo is less than 1.4, the whole spectral curve will be shifted downward by 0.6, while if it is greater than 1.4, it will be shifted downward by 0.7 (Figure \ref {Fig15}). The purpose of this shifted operation is to ensure that the spectral curves of these silicate-rich asteroids can be compared with the experiment spectra. Subject to slight space weathering, the slope of the mafic mineral spectra will increase slightly. The steep visible band region and overall inclined spectral curves reveal that these asteroids have suffered space weathering.

	%---------------------------Figures 15----------------------------
	%-----------------------------------------------------------------
	
\begin{figure}
		\centering
		\includegraphics[width=15.0cm, angle=0]{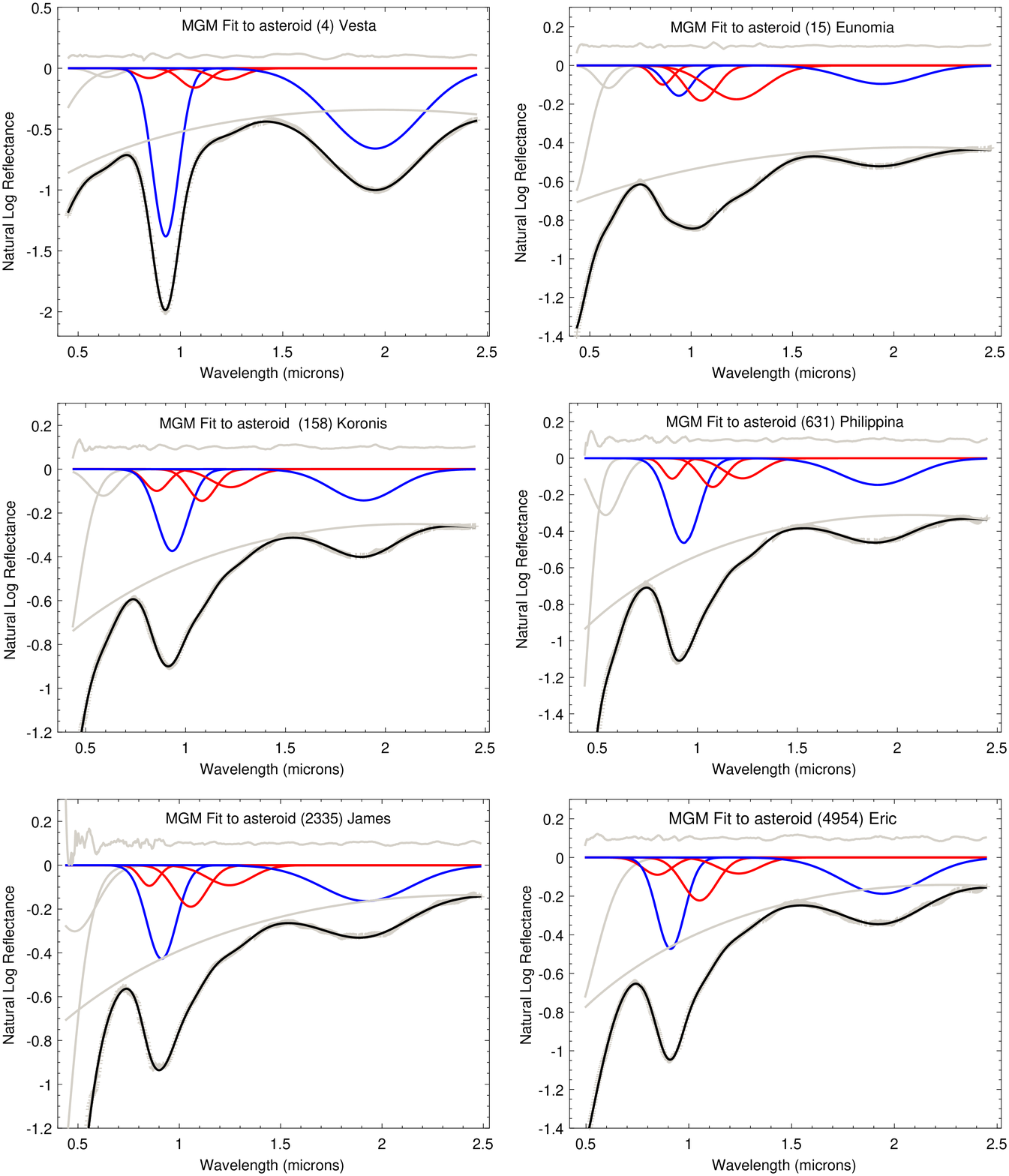}
		\caption{Decomposition of the components of 1.0 and 2.0 $\mu$m features for six asteroids (4) Vesta, (15) Eunomia, (158) Koronis, (631) Philippina, (2335) James and (4954) Eric. The OPX absorption bands fitted by MGM method are plotted as blue curves and red lines represent the absorption features with olivine. The black curves are the sum of all band profiles plus the continuum.}
		\label{Fig16}
	\end{figure}

	It is feasible to use the MGM deconvolution approach to estimate the mineral abundances. MGM with the logarithm of a second-order polynomial continuum removal method is employed to deconvolve these spectra. The deconvolution and estimated results are shown in Figure \ref {Fig16} and Table \ref {Tab5}. As shown in Figure \ref {Fig14} and described above, MGM will usually underestimate the proportion of olivine in the mixtures when deconvolution is applied to space weathered spectra, so the olivine content on these asteroids is only conservatively estimated in this paper. It means that the fresh rocks or subsurface materials of these asteroids may contain more olivine.

	%---------------------------Figures 16----------------------------
	%-----------------------------------------------------------------

	%------------------------Table  5----------------------------------
	\begin{table}
		\bc
		\begin{minipage}[]{150mm}
			\caption[]{MGM results and mineral abundance estimation results of the asteroids
				\label{Tab5}}\end{minipage}
		\setlength{\tabcolsep}{2.5mm}
		\small
		\begin{tabular}{lccccc}
			\hline\noalign{\smallskip}
			Asteroids&  Area\_1$\mu$m &Area\_2$\mu$m &OLV\_STR$_{{\rm M}1-2}$&Calculation &OLV : OPX\\
			\cline{1-6}
			(4) Vesta&	264.48&	341.82&	0.0939&	13.76&	33\% : 67\%\\
			(15) Eunomia&	127.96&	41.89&	0.1752&	1.87&	77\% : 23\%\\
			(158) Koronis&	123.33&	55.31&	0.0821&	5.46&	53\% : 47\%\\
			(631) Philippina&	144.59&	63.87&	0.1103&	4.00&	60\% : 40\%\\
			(2335) James&	142.63&	83.67&	0.0915&	6.41&	50\% : 50\%\\
			(4954) Eric&	152.56&	88.34&	0.0832&	6.96&	48\% : 52\%\\		
			\noalign{\smallskip}\hline
		\end{tabular}
		\ec
		\tablecomments{0.85\textwidth}{Calculation represents the value of MBAR/OLV\_STR$_{{\rm M}1-2}$}
	\end{table}
	%----------------------------------------------------------------

	\subsubsection{More Olivine on Asteroid (4) Vesta}
	\label{subsubsec:More Olivine on Asteroid (4) Vesta}

\begin{figure}
		\centering
		\includegraphics[width=15.0cm, angle=0]{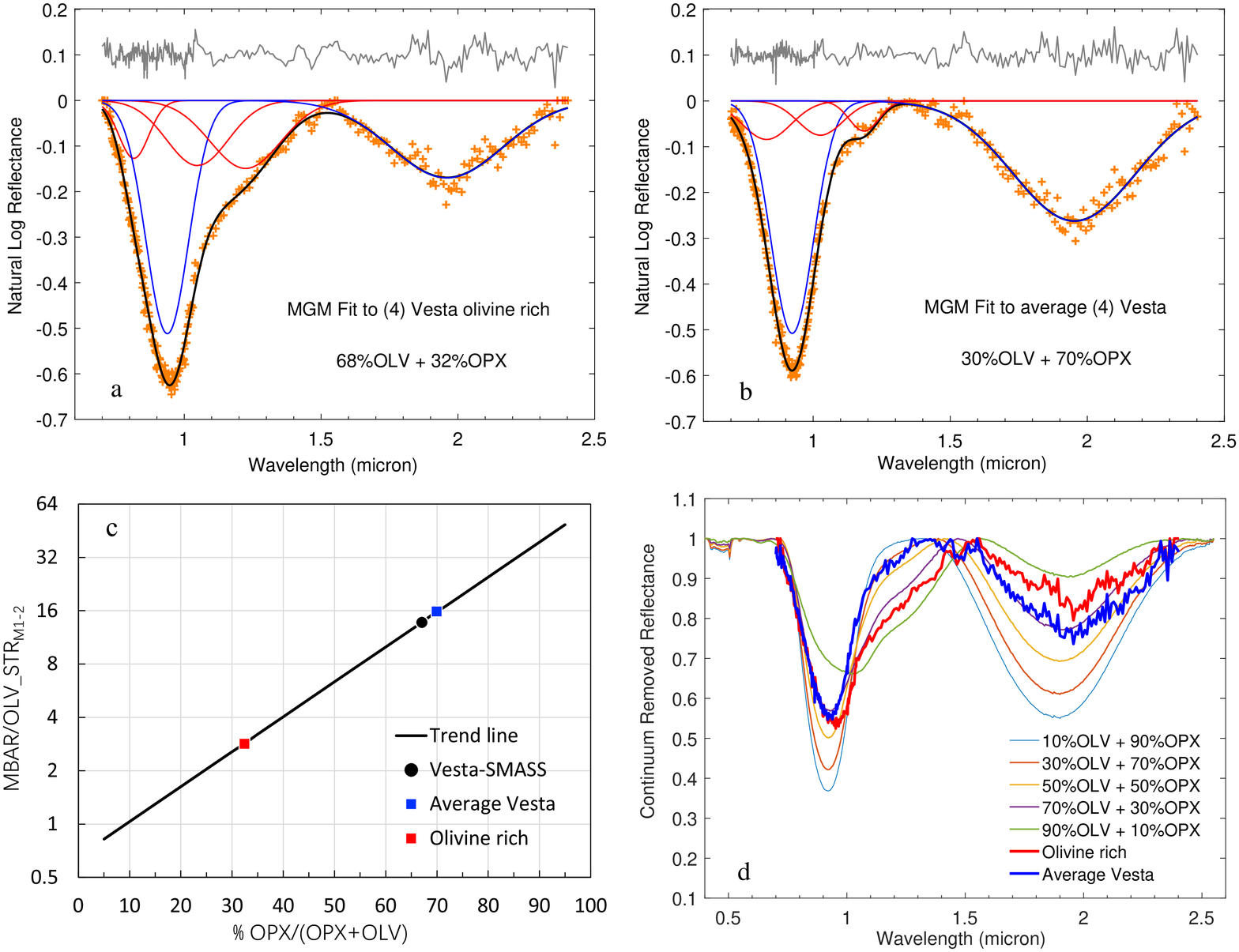}
		\caption{ Decomposition of the components of absorption features for main-belt asteroid (4) Vesta. (a). Modeled spectral fitting of the crater area with olivine-rich. (b). Modeled spectral fitting of average Vesta spectrum. (c). The estimations of mineral abundances derived from the SMASS and Dawn spectral data. (d). Comparison of the continuum-removed spectra obtained from various areas of (4) Vesta (Data from \cite{ammannito2013olivine} and olivine-orthopyroxene systematic mixtures with different proportions.}
		\label{Fig17}
	\end{figure}

	Asteroid (4) Vesta is the second most massive and probably the second largest asteroid (\citealt{thomas2005differentiation}). It has been considered that olivine and hypersthene are likely to be ubiquitous over the whole surface of Vesta (e.g., \citealt{hanna2009vesta, ammannito2013olivine, zambon2014spectral, le2015exploring, poulet2015modal}). \cite{ammannito2013olivine} inferred the presence of olivine using two primary methods, which are: Continuum removed spectral curve matching method and Band centers with the BAR parameters. The average spectra of (4) Vesta and olivine-rich site from a bright crater are given in that paper. The spectrum of the bright crater reveals clear olivine signatures, with the band near 1.0 $\mu$m centered at slightly longer wavelength than the average (4) Vesta spectrum, and the area of the 2.0 $\mu$m band is smaller than the average spectrum. The data are shown in Figure \ref {Fig17}d.

	%---------------------------Figures 17----------------------------
	%-----------------------------------------------------------------

	These data can infer that the surface of the asteroid (4) Vesta contains a high content of olivine. The olivine modeled abundance is about 20\%, and the spectra of some fresh rocks from the impact crater are interpreted as having an olivine content of more than 50\%, maybe even up to 70\% $\sim$ 80\% (\citealt{ammannito2013olivine,clenet2014deep, poulet2015modal}). In order to deconvolve these two continuum removed spectral data, we adjusted the MGM parameters to the study in this work. The deconvolution results (shown in Figure \ref {Fig17}) of the spectral data obtained by SMASS and Dawn are very consistent. Remote sensing and orbital average spectral data of the asteroid (4) Vesta show that there are about 30\% olivine + 70\% orthopyroxene in its surface mafic materials. While the deconvolution results of the bright crater spectrum reveals that there is approximately 70\% olivine. Considering that the spectraum of the bright impact crater is continuum removed (as described at the end of Section \ref {subsec:FRoIS}) and it has also been slightly weathered in space, we infer that the olivine composition of  partial craters on asteroid (4) Vesta are greater than 70\%. This result is consistent with the findings of \cite{ammannito2013olivine} and \cite{poulet2015modal}. Based on these findings, we can infer that the other asteroids mentioned above also have a high olivine content. However, space weathering prevents the correct deconvolution and interpretation of these spectra.

	\subsection{Different Orthopyroxenes}
	\label{subsec:Different Orthopyroxenes}
	
\begin{figure}
		\centering
		\includegraphics[width=15.0cm, angle=0]{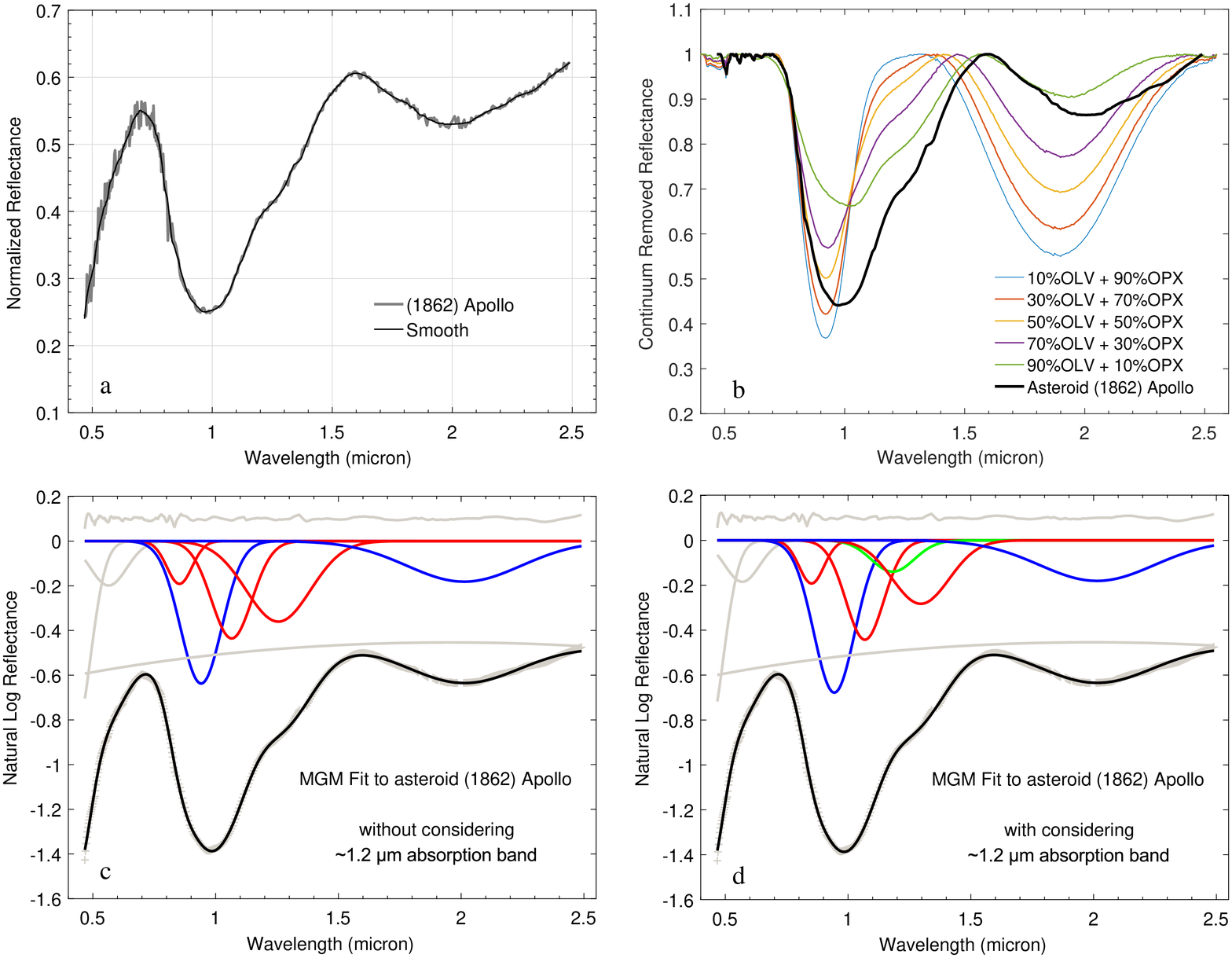}
		\caption{VIR Spectral curve of asteroid (1862) Apollo and  decomposition of the components of 1.0 and 2.0 $\mu$m features. (a). The observed and smoothed spectra are normalized at 0.55 $\mu$m and shifted downward to the suitable positions. (b). Comparison of the continuum-removed spectral curves and olivine-orthopyroxene mixtures with different proportions. (c). Decomposition of the components of absorption features without considering $\sim$1.2 $\mu$m sub-band composed of high-Fe orthopyroxene. (d). The curve is fitted by using same method as (c) panel, but considering $\sim$1.2 $\mu$m absorption band (marked as green line).}
		\label{Fig18}
	\end{figure}

	The orthopyroxene measured in RELAB and our experiment are both low-Fe orthopyroxenes. To understand whether the mixture of olivine and high-Fe orthopyroxene is also suitable for the method presented here, we carried out the following study. As is shown in Figure \ref {Fig1}b, the spectral curve of Ferrosillite, a kind of high-Fe orthopyroxene, shows an obvious absorption characteristic near 1.2 $\mu$m. This spin-allowed crystal field absorption band is attributed to the presence of Fe$^{2+}$ in the crystal M1 site, which is on the shoulder of the 1.0 $\mu$m band (\citealt{klima2007spectroscopy}). This position is very close to olivine's M1-2 band and thus care must be taken when interpreting the data. In addition, the two main absorption bands centered around 1.0 and 2.0 $\mu$m shift towards longer wavelength as the iron content increasing. Here we choose the asteroid (1862) Apollo (Figure \ref {Fig18}a) for further study, as its spectral curves show an olivine reflectance peak near 1.6 $\mu$m with two high-Fe orthopyroxene absorption bands (\citealt{gaffey1989reflectance, buratti20049969}).

	%---------------------------Figures 18----------------------------
	%-----------------------------------------------------------------

	Spectral curve matching method is used for this spectrum, and three absorption features near 1.0 $\mu$m, 1.2 $\mu$m and 2.0 $\mu$m are compared with systematic mixtures (Figure \ref {Fig18}b). It derived olivine abundance is about 80\% in the olivine-orthopyroxene mixtures. Two MGM deconvolution procedures are utilized here, one considering the $\sim$1.2 $\mu$m absorption of the high-Fe orthopyroxene and the other does not considering this band. The deconvolution results are shown in Figure \ref {Fig18}c and \ref {Fig18}d.

	If we ignore the absorption of the $\sim$1.2 $\mu$m band, the deconvolution result of MGM will get a larger absorption intensity in the olivine M1-2 band than the one who adds the $\sim$1.2 $\mu$m absorption band. The actual $\sim$1.2 $\mu$m absorption band will increase the area of 1.0 $\mu$m band, which will lead to the MBAR value decreasing. The combination of these two factors will be misinterpreted as the content of olivine in the material is higher than the actual abundance. Figure \ref {Fig18}c shows that the surface material of asteroid (1862) Apollo is composed of 96\% olivine and 4\% orthopyroxene, which does not meet our perception. However, Figure \ref {Fig18}d shows it contains about 88\% olivine and 12\% orthopyroxene (here we have deducted the area of the $\sim$1.2 $\mu$m absorption band). Although the results of these two estimated values are close, we have proved that the deconvolution result considering the $\sim$1.2 $\mu$m absorption band is relatively reliable.
	
	\section{Conclusions}
	\label{sect:Conclusions}
	
	In this article, a new empirical procedure based on MGM with some prior mineral absorption bands is presented to estimate the end-members abundance from olivine-orthopyroxene mixture spectra. The accuracy of the end-member abundance can be estimated within 15\% by using this new procedure.

	Integrated with RELAB database, a new spectral set obtained from  olivine-orthopyroxene mixtures assembled at 20\% mineral abundance intervals in our experiments is exploited. The VNIR  absorption spectra of olivine-orthopyroxene mixtures systematically varied with their various abundances. However, we found the commonly-used BAR does not work well for some mixtures and the BAR of different mixture sets show large deviations. As a powerful tool, MGM is utilized to reveal the composition of mafic mineral assemblages. A modified BAR (namely MBAR) is introduced to describe new-defined band area ratio based on the fitted parameters by MGM. Two parameters of MBAR and olivine M1-2 band absorption strength (as defined in the MGM) are used to estimate the mineral abundances of olivine-orthopyroxene mixtures. The systematic procedure is quite efficient for deconvolution of olivine-orthopyroxene mixture spectra and estimation of the mineral abundances for mafic materials from the VNIR reflectance spectra.

	Additionally, space weathering effect on olivine-orthopyroxene mixtures and silicate asteroids is investigated based on the new proposed procedure. Some samples are irradiated by high-energy laser in our experiments and the spectral profiles of these samples vary with irradiation fluxes significantly. After irradiation, we find that mineral diagnostic spectral features ($\sim$1.0 and $\sim$2.0 $\mu$m bands) become weaker and shallower. The intensities of absorption bands and albedos of these spectra in the visible region decrease faster than that in the NIR region. The MGM deconvolution results of these irradiated mixtures indicate that orthopyroxene has stronger resistance to space weathering than olivine. Among the results, MBAR increases, while olivine M1-2 band absorption strength  decreases, which makes these irradiated spectra more difficult to identify and interpret. This usually leads to an underestimate of olivine abundance. The VNIR spectra of main-belt asteroid (4) Vesta from SMASS catalog and Dawn mission are investigated by applying the new procedure and the results are consistent with the others' studies.

	Besides, the influence of absorption band strength of high-Fe orthopyroxene at $\sim$1.2 $\mu$m is also discussed in this study and we claim that this band should not be ignored in the analysis of olivine and high-Fe orthopyroxene mixtures. In the future, more experiments for spectral deconvolution of different mafic mixtures and studies for various space weathering effects will be implemented to extend the applications of the new proposed procedure and MGM.

	\normalem
	
	\begin{acknowledgements}
		
		We thank the anonymous reviewer for useful comments to improve our manuscript. This work is supported by the Foundation of the State Key Laboratory of Lunar and Planetary Sciences, Macau University of Science and Technology, Macau, China. X. Lu is also funded by The Science and Technology Development Fund, Macau SAR (No. 0073/2019/A2). C. H. Hsia acknowledges the support from The Science and Technology Development Fund, Macau SAR (No. 0007/2019/A). Y. Yang is supported by Beijing Municipal Science and Technology Commission (No. Z181100002918003). T. Jiang and H. Zhang are supported by Natural Science Foundation of China (Nos. 11773023, 11941001, 12073024 and U1631124).  We are very grateful to Xiao-Yi Hu and Pei Ma for their assistances in the experiments, whose comments and suggestions also improved the manuscript. Part of the spectral data utilized in this study are obtained and make available by the RELAB Spectral Database at Brown University and Planetary Spectroscopy data at MIT. The MGM program is developed and provided by Brown University. We acknowledge the support of Brown University and MIT.
		
	\end{acknowledgements}

	%\bibliographystyle{raa}
	%\bibliography{HHJBIB}

\end{document}